\documentclass[12pt]{article}

\usepackage{amsmath}
\usepackage{amssymb}
\usepackage{delarray}
\usepackage{graphics}
\usepackage{color}
\usepackage{axodraw}
\input epsf
\providecommand{\SU}[1]{}
\renewcommand{\SU}[1]{\ensuremath{\mathrm{SU}(#1)}}
\providecommand{\UI}{}
\renewcommand{\UI}{\ensuremath{\mathrm{U}(1)}}
\providecommand{\psgroup}{}
\renewcommand{\psgroup}{\ensuremath{\mathrm{SU}(4) \otimes \mathrm{SU}(2)_L \otimes \mathrm{SU}(2)_R } }
\providecommand{\smgroup}{}
\renewcommand{\smgroup}{\ensuremath{\mathrm{SU}(3) \otimes \mathrm{SU}(2)_L \otimes \mathrm{U}(1)_Y } }

\providecommand{\Dpbranes}{}
\renewcommand{\Dpbranes}{\ensuremath{Dp\mathrm{-branes}}}
\providecommand{\Dbrane}[1]{}
\renewcommand{\Dbrane}[1]{\ensuremath{D#1\mathrm{-brane}}}
\providecommand{\Dbranes}[1]{}
\renewcommand{\Dbranes}[1]{\ensuremath{D#1\mathrm{-branes}}}
\providecommand{\vev}{}
\renewcommand{\vev}{vev}
\providecommand{\vevs}{}
\renewcommand{\vevs}{vevs}

\newcommand{\sheptitle}
{\Large Lepton flavour violation in realistic non-minimal supergravity models}

\newcommand{\shepauthor}
{J.~G.~Hayes, S.~F.~King, I.~N.~R.~Peddie \footnote{Address after 
1st September, 2005: Physics Division, School of Technology,
Aristotle University of Thessaloniki, Thessaloniki 54124, Greece.}}

\newcommand{\shepaddress}
{School of Physics and Astronomy, University of Southampton, \\
        Southampton, SO17 1BJ, U.K.}

\newcommand{\shepabstract} {Realistic effective supergravity models
have a variety of sources of lepton flavour violation (LFV) which can
drastically affect the predictions relative to the scenarios
usually considered in the literature based on minimal supergravity and
the supersymmetric see-saw mechanism.
We catalogue the additional sources of LFV which occur in realistic
supergravity models including the effect of D-terms arising from an Abelian
$U(1)$ family symmetry, non-aligned trilinear contributions from
scalar F-terms, as well as non-minimal supergravity contributions
and the effect of different Yukawa textures.  
In order to quantify these effects, we investigate a string inspired
effective supergravity model arising from intersecting D-branes 
supplemented by an additional $U(1)$ family symmetry. In such 
theories the magnitude of the D-terms is predicted, and we 
calculate the branching ratios for 
$\mu\rightarrow e\gamma$ and $\tau \rightarrow \mu \gamma$ 
for different benchmark points designed to isolate the 
different non-minimal contributions. 
We find that the D-term
contributions are generally dangerously large, but in certain cases
such contributions can lead to a dramatic suppression of LFV rates,
for example by cancelling the effect of the see-saw induced
LFV in $\tau \to \mu \gamma$ models with lop-sided textures.
In the class of string models considered here we find the surprising result
that the D-terms can sometimes serve to restore 
universality in the effective non-minimal supergravity theory.}

\begin{document}


\begin{titlepage}


\begin{center}
{\large{\bf \sheptitle}} \\
\vspace{0.5in}
\bigskip \shepauthor \\ \mbox{} \\ {\it \shepaddress} \\
\vspace{0.5in}
\bigskip {\bf Abstract} \bigskip
\end{center}
\setcounter{page}{0}
\shepabstract
\begin{flushleft}
\today
\end{flushleft}

\vskip 0.1in
\noindent


\end{titlepage}



\newpage 
\section{Introduction} 
\label{sec:introduction}



Lepton flavour violation (LFV) is a sensitive probe of new physics
in supersymmetric (SUSY) models
\cite{Borzumati:1986qx,Gabbiani:1996hi}.
In SUSY models, LFV arises due to off-diagonal elements in the
slepton mass matrices in the `super-CKM' (SCKM) basis, in which the
quark and lepton mass matrices are diagonal \cite{Chung:2003fi}.
In supergravity (SUGRA) mediated
SUSY breaking the soft SUSY breaking masses are generated at the
Planck scale, and the low energy soft masses relevant for 
physical proceses such as LFV are therefore
subject to radiative corrections in running from the Planck scale
to the weak scale. Off-diagonal slepton masses in the SCKM basis
can arise both directly at the high energy scale
(due to the effective SUGRA theory which is responsible for them), or can
be radiatively generated by renormalisation group running
from the Planck scale to the weak scale, 
for example due to Higgs triplets in GUTs or right-handed neutrinos
in see-saw models which have masses intermediate between these
two scales.

Neutrino experiments
confirming the Large Mixing Angle (LMA) MSW solution to the
solar neutrino problem \cite{GoPe} taken together with the
atmospheric neutrino data \cite{SKamiokandeColl} shows that neutrino
masses are inevitable \cite{King:2003jb}.
The presence of right-handed neutrinos, as required by the see-saw
mechanism for generating neutrino masses, will lead inevitably to LFV,
due to running effects, even in minimal SUGRA 
(mSUGRA) which has no LFV at the high energy
GUT or Planck scale \cite{Hisano:1995cp,King:1998nv}. 
Therefore, merely assuming SUSY and the see-saw
mechanism, one expects LFV to be present. This has been studied, for
example, in mSUGRA models with a natural neutrino mass hierarchy
\cite{Blazek:2002wq}. There is a large literature on 
the case of minimal LFV arising from
mSUGRA and the see-saw mechanism \cite{huge}.


Despite the fact that most realistic string models lead
to a low energy effective {\em non-minimal} SUGRA theory,
such theories have not been extensively studied in the literature,
although a general analysis of flavour changing effects
in the mass insertion approximation has recently been performed
\cite{Chankowski:2005jh}. Such effective non-minimal SUGRA models 
predict non-universality of the soft masses at the high energy scale,
dependent on the structure of the Yukawa matrices. Moreover there can be
additional sources of LFV which also enter the analysis.
For example, realistic effective SUGRA theories arising from
string-inspired models will also typically involve
some gauged family symmetry which can give an additional
direct (as opposed to renormalisation group
induced) source of LFV. This is because the D-term
contribution to the scalar masses generated when the family symmetry
spontaneously breaks adds different diagonal elements 
to each generation in the `theory' basis, and generates
non-zero off-diagonal elements in the SCKM basis, leading to LFV. This effect
depends on the strength of the D-term contribution, which is expected
to be close in size to the size of the uncorrected scalar masses.
There can also be a significant contribution to
non-universal trilinear soft masses leading to flavour violation
\cite{Abel:2001ur,Ross:2002mr,KP0307091} arising from the 
F-terms of scalars associated with the Yukawa couplings
(for example the flavons of Froggatt-Nielsen theories). 

The purpose of the present
paper is to catalogue and quantitatively study the importance of
all the different sources of LFV present
in a general non-minimal SUGRA framework, including the 
effects of gauged family symmetry.
Although the different effects have all been identified in the literature,
there has not so far been a coherent and quantitative
dedicated study of LFV processes, beyond the mass insertion approximation,
which includes all these effects within a single framework.
In order to quantify the importance of the different effects it is necessary
to investigate these disparate souces of LFV numerically, both in isolation and
in association with one another, within some particular SUGRA 
model. To be concrete we shall study the effective 
SUGRA models of the kind considered
in \cite{KP0307091} which have a sufficiently rich structure
to enable all of the effects to be studied within a single framework.
Within this class of models we shall consider specific benchmark
points in order to illustrate the different effects.
Some of these benchmark points were already previously considered
in \cite{KP0307091}. However in the previous study the important effect
of D-terms arising from the Abelian family symmetry was not
considered. Here we shall show that such D-terms are in fact
calculable within the framework of the particular model
considered, and can lead to significant enhancement
(or suppression) of LFV rates, depending on the particular 
model considered. 

The outline of this paper is as follows. In
Section~\ref{sec:soft-from-SUGRA} we discuss soft
supersymmetry breaking masses in supergravity. In
Section~\ref{sec:sourc-lfv} we summarise the flavour problem and
catalogue the distinct sources of lepton flavour violation. In
Section~\ref{sec:442-pati-salam}, we outline the aspects of
the specific models that we shall study. In Section~\ref{soft} we
discuss the soft SUSY-breaking sector within this class of models, 
parameterise the F-terms, and write down the soft terms, including the D-term
contribution to the scalar masses. We also discuss the D-terms
associated with a family symmetry that are expected to lead to large
lepton flavour violation. In Section~\ref{sec:numerical-etc} we
specify two models in this class, define benchmark points
and present the results of our numerical analysis of LFV for these
benchmark points.
Section~\ref{sec:conclusions} concludes the paper.


\section{Soft terms from supergravity}
\label{sec:soft-from-SUGRA}


We summarise here the standard way of getting soft SUSY breaking terms
from supergravity.  Supergravity is defined in terms of a K\"ahler
function, $G$, of chiral superfields ($\phi = h, C_a$). Taking the
view that supergravity arises as the low energy effective field theory
limit of a string theory, the hidden sector fields $h$ are taken to
correspond to closed string moduli states ($h = S, T_i$), and the
matter states $C_a$ are taken to correspond to open string states. In
string theory, the ends of the open string states are believed to be
constrained to lie on extended solitonic objects called \Dpbranes.

Using natural units,
\begin{equation}
  \label{eq:Kahler_function}
  G(\phi, \overline{\phi}) =
  \frac{K(\phi,\overline{\phi})}{\tilde{M}^2_{Pl}}
  + \ln
  \left(
    \frac{W(\phi)}{\tilde{M}^3_{Pl}}
  \right)
  + \ln
  \left(
    \frac{W^*(\overline{\phi})}{\tilde{M}^3_{Pl}}
  \right)
\end{equation}

$K(\phi,\overline{\phi})$ is the K\"ahler potential, a real function of
chiral superfields. This may be expaned in powers of $C_a$:
\begin{equation}
  \label{eq:Kahler_potential}
  K = 
  \overline{K}(h,\overline{h})
  + \tilde{K}_{\overline{a}{b}}(h,\overline{h}) \overline{C}_{\overline{a}} C_b
  + 
  \left[
    \frac{1}{2}Z_{ab}(h,\overline{h})C_a C_b + h.c. \right]
  + ...
\end{equation}

$\tilde{K}_{\overline{a} b}$ is the K\"ahler metric. $W(\phi)$ is the
superpotential, a holomorphic function of chiral superfields.
\begin{equation}
  \label{eq:superpotential}
  W = \hat{W}(h) + \frac{1}{2}\mu_{ab}(h)C_a C_b +
  \frac{1}{6}Y_{abc}C_a C_b C_c + ...
\end{equation}

We expect the supersymmetry to be broken; if it is broken, then the
auxilliary fields $F_\phi \ne 0$ for some $\phi$. As we lack a model of
SUSY breaking, we introduce Goldstino angles as parameters that will
enable us to explore different methods of breaking supersymmetry. We
introduce a matrix, $P$ that canonically normalises the K\"ahler metric,
$P^\dagger K_{\overline{J}I} P = 1$
\footnote{ The subscripts on the K\"ahler potential $K_I$ means
$\partial_I K$. However, the subscripts on the F-terms are just
labels. } \cite{P_Matrix:introduction}.  We also introduce a column
vector $\Theta$ which satisfies $\Theta^\dagger\Theta = 1$. We are
completely free to parameterise $\Theta$ in any way which satisfies
this constraint.

Then the un-normalised soft terms and trilinears appear in the soft
SUGRA breaking potential \cite{Brignole:1997dp}
\begin{equation}
  \label{eq:soft_potential}
  V_{\mathrm{soft}} = m^2_{\overline{a}b} \overline{C}_{\overline{a}} C_b
    + \left( \frac{1}{6}A_{abc} Y_{abc} C_a C_b C_c + h.c. \right) + ...
\end{equation}

The non-canonically normalised soft trilinears are then
\begin{eqnarray}
  \nonumber
  A_{abc} Y_{abc}
  &=&
  \frac{\displaystyle \hat{W}^*}{|\displaystyle \hat{W}| }
  e^{\overline{K}/2} F_m
  \left[
    \overline{K}_m Y_{abc} + \partial_m Y_{abc} - 
    \left(
      \left(\tilde K^{-1}\right)
    \right. 
  \right. \partial_m
  \tilde K_{\overline{e}a} Y_{dbc}
  \\
  \label{eq:unnomalised_trilinears}  
  && 
  \left. 
    \left.
      { \color{white} \tilde K^-1_a }
      {} + ( a \leftrightarrow b ) 
      + ( a \leftrightarrow c )
    \right)
  \right]
\end{eqnarray}

In this equation, it should be noted that the index $m$ runs over $h,
C$. However, by definition, the hidden sector part of the K\"ahler
potential and the K\"ahler metrics is independent of the matter
fields.

Assuming that the terms $\partial_C Y_{abc} \ne 0$,
the canonically normalised equation for the trilinear is
\begin{equation}
  \label{eq:normalised_trilinear}
  A_{abc} = F_I 
  \left[
    \overline{K}_I - \partial_I \ln
    \left(
      \tilde K_a \tilde K_b \tilde K_c
    \right)
  \right]
  + F_m \partial_m \ln Y_{abc}
\end{equation}
If the Yukawa hierarchy is taken to be generated by a Froggatt-Nielsen
(FN) field, $\phi$, such that $Y \propto \phi^p$, then we expect $F_\phi
\propto m_{3/2} \phi$, and then $F_\phi \partial_\phi \ln Y \propto
m_{3/2}$, and so even though these fields are expected to have heavily
sub-dominant F-terms, they contribute to the trilinears on an equal
footing as the moduli.

If the K\"ahler metric is diagonal and non-canonical, then the
canonically normalised scalar mass-squareds are given by
\begin{equation}
  \label{eq:normalised_scalars}
  m_a^2 = m^2_{3/2} - F_{\overline{J}} F_I \partial_{\overline{J}} \partial_I 
  \left( \ln \tilde{K_a} \right) ,
\end{equation}

and the gaugino masses are given by
\begin{equation}
  \label{eq:normalised_gauginos}
  M_\alpha = \frac{1}{2 \mathrm{Re} f_\alpha } F_I \partial_I f_\alpha ,
\end{equation}
where $f_\alpha$ is the `gauge kinetic function'. $\alpha$ enumerates
$D$-branes in the model.  In type~I string models without twisted
moduli these have the form $f_9 = S\; , \; f_{5_i} = T_i$.

Specifically, we use a K\"ahler potential that doesn't have any
twisted-moduli \cite{Ibanez:1998rf}:
\begin{eqnarray}
  \nonumber K &=& -\ln \left( S + \overline{S} -
  \left|C_1^{5_1}\right|^2 - \left|C_2^{5_2}\right|^2 \right) -\ln
  \left( T_1 + \overline{T}_1 - \left|C^9_1\right|^2 -
  \left|C_3^{5_3}\right|^2 \right) \\ \nonumber && {} - \ln \left( T_2
  - \overline{T}_2 - \left|C_2^9\right|^2 - \left|C_3^{5_1}\right|^2
  \right) -\ln \left( T_3 - \overline{T}_3 -\left|C_3^9\right|^2 -
  \left|C_2^{5_1}\right|^2 - \left|C_1^{5_1}\right|^2 \right) \\
  \nonumber && {} + \frac{\left|C^{5_1 5_2}\right|^2
  }{\left(S+\overline{S}\right)^{1/2}\left(T_3+\overline{T}_3\right)^{1/2}}
  + \frac{\left|C^{95_1}\right|^2}{\left(T_2 +
  \overline{T}_2\right)^{1/2}\left(T_3 + \overline{T}_3\right)^{1/2}}
  \\
  \label{eq:62}
  && {} +
  \frac{\left|C^{95_2}\right|^2}{\left(T_1+\overline{T}_1\right)^{1/2}
  \left(T_3+\overline{T}_3\right)^{1/2}}
\end{eqnarray}

The notation is that the field theory scalars, the dilaton $S$ and the
untwisted moduli $T_i$ originate from closed strings. Open string
states $C^b_i$ are required to have their ends localised onto
\Dbranes{}. The upper index then specifies which brane(s) their ends
are located on, and if both ends are on the same brane, the lower
index specifies which pair of compacitified extra dimensions the
string is free to vibrate in.


\section{Sources of lepton flavour violation}
\label{sec:sourc-lfv}


There are two parts to the flavour problem. The first is understanding
the origin of the Yukawa couplings (and heavy Majorana masses for the
see-saw mechanism), which lead to low energy quark and lepton mixing
angles. In low energy SUSY, we also need to understand why flavour
changing and/or CP violating processes induced by SUSY loops are so small.
A theory of flavour must address both problems simultaneously. For a
full discussion of this see the review \cite{Chung:2003fi}.

There are two contributions that can lead to large amounts of flavour
violation. The first is the non-alignment of the trilinear soft
coupling matrices to the corresponding Yukawa matrices, due to the
contribution $F_m \partial_m Y$, $m = \{ H, \overline{H}, \theta,
\overline\theta\}$.  The reasons why this can lead to large flavour
violation are have been given before \cite{KP0307091}, where a
numerical investigation of a model very similar to those considered
herein finds that there is a large amount of flavour violation.
The second contribution can come from scalar mass matrices which are
not proportional to the identity in the theory basis, and lead to 
off-diagonal entries in the SCKM basis, resulting in flavour violation.

In this section we begin by defining the SCKM basis, and 
in the following subsections we systematically
discuss a number of distinct sources of Lepton Flavour Violation (LFV) in 
SUGRA models. As well as considering generic SUGRA models, we also allow for a
family symmetry, which easily lead to non-universal scalar mass matrices, and
non-aligned trilinear matrices\footnote{By non-aligned trilinears, we mean that
$\tilde{A}_{ij}/Y_{ij} \ne \mathrm{constant} \ \ (\mathrm{no\ sum}).$
}.


\subsection{The SCKM Basis}
\label{sec:high_fv}


The most
convenient basis to work in for considering flavour violating decays,
such as $\mu\rightarrow e\gamma$ is the super-CKM (SCKM) basis, which is the
basis where the Yukawa matrices are diagonal. If we define the unitary
rotation matrices $U_f, V_f$ by
\begin{equation}
  \label{eq:3} Y^f_{\mathrm{diag}} = U_f Y^f V^\dag_f ,
\end{equation} such that $Y^f_{\mathrm{diag}}$ has positive
eigenvalues. To convert the physical mass matrices to the SCKM basis,
we rotate by the relevant matrix; $U_f$ for the left-handed scalar
matrices and $V_f$ for the right-handed. Then flavour violation is
proportional to the off-diagonal elements in the SCKM basis, and is
suppressed by the diagonal values. The selectron mass matrix is 6 by
6, and the sneutrino mass matrix is 3 by 3 \footnote{The heavy right
handed neutrinos cause the right-handed part of the sneutrino mass
matrix to decouple by the electroweak scale.}. The selectron mass
matrix is
\begin{equation}
  \label{eq:5}
  m^2_{\tilde E} =
  \left(
    \begin{array}{cc}
      Y^e {Y^e}^\dag v_1^2 + \frac{1}{4}v^2(g_2^2 - g_1^2)\mathbf{1} + m_{LL}^2 &
      -Y^e v_2 \mu + \tilde{A^e} v_1 \\
      {-Y^e}^\dag v_2 \mu^* + \tilde{A^e}^\dag v_1 &
      {Y^e}^\dag Y^e v_1^2 + \frac{1}{2}v^2 g_1^2 \mathbf{1}  + m_{ER}^2 
    \end{array}
  \right) ,
\end{equation}
where $v^2 = v_2^2 - v_1^2$. The sneutrino mass matrix is then
\begin{equation}
  \label{eq:6}
  m^2_{\tilde \nu} =
  \left(
    Y^\nu {Y^\nu}^\dag v_2^2 + \frac{1}{4} v^2 g_1^2\mathbf{1} + m_{LL}^2
  \right)
\end{equation}

Off diagonal elements in any of the 3 by 3 submatrices in the SCKM
basis will lead to flavour violation.  We will now consider the $LL$
block of $m^2_{\tilde{E}}$. The arguments follow for any other block
of $m^2_{\tilde{E}}$ or $m^2_{\tilde \nu}$. The transformation to the
SCKM basis is carried out by
\begin{equation}
  \label{eq:7}
  m^2_{\tilde E,LL} = U_e \left( Y^e {Y^e}^\dag v_1^2 +
\frac{1}{4}v^2(g_2^2 - g_1^2)\mathbf{1} + m_{LL}^2 \right) U_e^\dag  
\end{equation}

$U_e$ is unitary, and $U_e Y^e V_e^\dag$ is diagonal, so the first two
terms will be diagonal. Any off-diagonality must come from the third
term. If this is proportional to the identity at the GUT scale, it
will be approximately equal to the identity at the electroweak scale,
which is the scale we should be working at. The fact that this is only
approximate is due to the presence of the right handed neutrino fields
in the running of the soft scalar mass squared matrices. If, however,
the soft mass squared matrices are not proportional to the identity at
the GUT scale, then large off-diagonal values will be generated when
rotating to the SCKM basis, unless the rotation happens to be
small. Generally this won't be the case. Since the family D-term
contribution is not proportional to the identity\footnote{This
statement assumes that the generational charges are not the same for
both left- and right-handed fields. This would remove the point of
the family symmetry generating the fermion mass hierarchy.} this will
usually be the case\footnote{One can, however, imagine some model with
aberrant points in its parameter space where a non-universal non-zero D-term corrects a
non-universal base mass matrix to give a universal net mass matrix.}
and so we expect large flavour violation in models with Abelian family
symmetries when the D-terms correct the scalar mass matrices.


\subsection{The relevance of the Yukawa textures}
\label{sec:relev-yukawa-textures}


There is one subtlety concerning the size of the off-diagonal elements of the
scalar mass matrices in the SCKM basis.  This comes back to the
definition of the SCKM basis as the basis in which the Yukawa matrices
are diagonal. The larger the SCKM transformation between any `theory'
basis and the mass eigenstate basis for the Yukawa matrices, the
larger the SCKM transformation that must be performed on the 
scalar mass matrices in going to the SCKM basis,
hence the larger the off-diagonal elements of the 
scalar mass matrices in the SCKM basis generated from non-equal
diagonal elements in the `theory' basis.
\footnote{
Note that the D-terms make us sensitive to right-handed mixings in the Yukawa
matrices, so the non-universal family charge structure for the right-handed
scalar masses may lead to a non-universal generational hierarchy in
the right-handed scalar mass matrices.}
The larger the off-diagonal entries in the SCKM basis
compared to the diagonal ones, the greater will be 
the flavour violation. Also, the greater the mass difference between the
diagonal elements in the `theory' basis, 
the greater the size of the off-diagonal entries
produced when rotating from the 'theory' basis, hence the larger the
flavour violating effect. Clearly these effects are sensitive to the 
size of the transformation required to go to the SCKM basis,
which in turn is sensitive to the particular choice of Yukawa
textures in the `theory' basis. In this way, the choice of Yukawa 
texture can play an important part on controlling the magnitude
of flavour violation, and we shall see examples of this later.


\subsection{Running effects}
\label{sec:running-effects-only}


Consider the case where, at the high-energy scale, the scalar mass
matrices are proportional to the identity matrix and each soft
trilinear coupling matrix is aligned to the corresponding Yukawa
matrix:
\begin{equation}
  \label{eq:8} \left(m^2_{f}\right)_{ij} = m^2_{0,f} \delta_{ij}
\;\;\;,\;\;\; \left(\tilde{A}^f\right)_{ij} = A_0^f Y^f_{ij} .
\end{equation}
This is often referred to as mSUGRA.
In the quark sector, due to the quark flavour violation responsible
for CKM mixing, 
when the scalar squark 
mass matrices are run down to the electroweak scale, they
will run to non-universal scalar mass matrices and non-aligned
trilinear coupling matrices. If this is the case, then in the SCKM
basis, which is the basis where the Yukawa matrices are diagonal,
off-diagonal elements in the scalar squark mass matrices or the trilinear
squark mass matrices lead to flavour violation.

In the lepton sector, in the absence of neutrino masses
the separate lepton flavour numbers are conserved and mSUGRA will not lead to
any LFV induced by running the matrices down to low energy. 
However, in the presence of neutrino masses, 
with right-handed neutrino fields included 
to allow a see-saw explanation of neutrino masses and
mixing angles, the separate lepton flavour numbers will be violated and, 
even in the mSUGRA type scenario, running effects
will generate off-diagonal elements in the scalar
mass matrices in the SCKM basis, resulting in low energy LFV.





\subsection{Diagonal scalar mass matrices not propotional to the unit matrix}
\label{sec:diagonal-scalar-mass}



\subsubsection{Non-minimal SUGRA}
\label{sec:non-minimal-sugra-dsm}


In non-minimal SUGRA the scalar mass matrices may be diagonal at the
high-energy scale, but not proportional to the identity. In this case,
there will be non-zero off-diagonal elements in the SCKM basis even
with no contribution to running effects, or contribution from the
trilinear coupling matrices.

One way of getting diagonal mass matrices not proportional to the
unit matrix is from a SUGRA model
corresponding to the low energy limit of a string model with
D-branes. If each generation from the field theory viewpoint
corresponds to a string attaching to different branes, then the masses
predicted in the SUGRA can be different. This leads to diagonal but
non-universal scalar mass matrices.


\subsubsection{D-term contributions from broken family gauge groups}
\label{sec:d-term-contributions-sbfgg}


Another way of getting diagonal mass matrices not proportional
to the unit matrix is by having a model
with a gauge family symmetry, which is broken spontaneously. When the
Higgs which breaks the family group, the flavon, gets a vev, it
contributes a squark (slepton) mass contribution through the four
point scalar gauge interaction which has two flavons and two  squarks
(sleptons).

To make the point more explicitly, consider a $U(1)$ family
group. Then the mass contribution is proportional to the charge under
the family symmetry. As the point of a family symmetry is to explain
the hierarchy of fermion masses, small quark mixing angles and large
neutrino mixing angles, the charges are usually different.

Then, even if the mass matrix starts off as a universal matrix, it
will be driven non-universal by the D-term contribution:
\begin{equation}
  \label{eq:9} m^2_{L_L} = \left[
    \begin{array}{ccc} m_0^2 \\ & m_0^2 \\ && m_0^2
    \end{array} \right] + D^2 \left[
      \begin{array}{ccc} q_{L1} \\ & q_{L2}  \\ && q_{L3}
      \end{array} \right] .
\end{equation}


\subsection{Non-aligned trilinears}
\label{sec:non-aligned trilinears}



\subsubsection{Non-minimal SUGRA}
\label{sec:non-minimal SUGRA}

One way of getting non-aligned trilinear matrices is by having the
same sort of non-minimal SUGRA setup that leads to diagonal but
non-universal mass matrices, as described in Section
\ref{sec:non-minimal-sugra-dsm}. From the supegravity equations from
Section \ref{sec:soft-from-SUGRA}, the trilinears that appear in the
soft Lagrangian, $\tilde{A}_{ij}$ will be non-aligned if the
trilinears predicted by the SUGRA model, $A_{ij}$ are not democratic,
i.e. if $A_{ij} \ne \mathrm{constant}$.  From a string-inspired/SUGRA
standpoint, if each generation is assigned to a different brane and
extra-dimensional vibrational direction, then in general we expect
$A_{abc}$ to be different, due to the differing values of the K\"ahler
metrics $\tilde{K}_a$ for the different brane assignents $C^i_j$.
When $\tilde{A}_{ij}$ is transformed to the SCKM basis at the
electroweak scale, there will then be large off-diagonal elements
which contribute to flavour violating processes.

\subsubsection{Flavon contributions from the Yukawa couplings}
\label{sec:FNcontributions}

In general 
when one considers a family symmetry in order to understand the origin
of the Yukawa couplings, the new fields arising
from this can develop F-term \vev{}s, and contribute to the supersymmetry
breaking F-terms in a non-universal way. This leads to a dangerous source
of flavour violating non-aligned trilinears:
\cite{Abel:2001ur,Ross:2002mr,KP0307091,King:2003xn},
\begin{equation}
  \label{eq:10}
  \Delta A  = F_\theta \partial_\theta \ln Y 
\end{equation}
where the Yukawa coupling $Y$ in Eq.~(\ref{eq:10}) arises from the
an effective FN operator and is a polynomial of the FN field $\theta$,
$Y \sim \theta^n $, leading to
\begin{equation}
\Delta A = F_\theta \partial_\theta \mathrm{ln} \theta^n = F_\theta
\frac{n}{\theta} .
\end{equation}
However the auxiliary field is proportional to the scalar component,
\begin{equation}
F_\theta \propto m_{3/2} \theta \Rightarrow \Delta A \propto nm_{3/2} .
\end{equation}
An example of this with an arbritary $U(1)$ family symmetry is
\begin{equation}
Y_{ij} = a_{ij} \left( \frac{\theta}{M} \right)^{p(i,j)} \Longrightarrow 
\Delta A_{ij} \sim m_{3/2} p(i,j)
\end{equation}
The $a_{ij}$ are arbiratry couplings, all of which should be $O(1)$
for the symmetry to be considered natural. The $p(i,j)$ are
integers appearing as a power for the $ij$-th element of the above
Yukawa, and it comprises the sum of the family charges for the
$i$th-generation left-handed field and $j$th-generation right-handed
field.
In principle, if the Yukawa texture is set up so that each power is different,
then each element in $A_{ij}$ will be different from each other, and the
physical trilinear matrix, $\tilde{A}_{ij}$ will be non-aligned to the
corresponding Yukawa. 
Due to the dependence on the charges of the different fields,
this contribution to the trilinears is not diagonalised when we transform
to the SCKM basis.

\section{Intersecting D-brane models with an Abelian family symmetry}
\label{sec:442-pati-salam}


\subsection{Symmetries and symmetry breaking}
\label{sec:symm and symm breaking}


\begin{figure}
\epsfysize=2.5truein
\epsffile{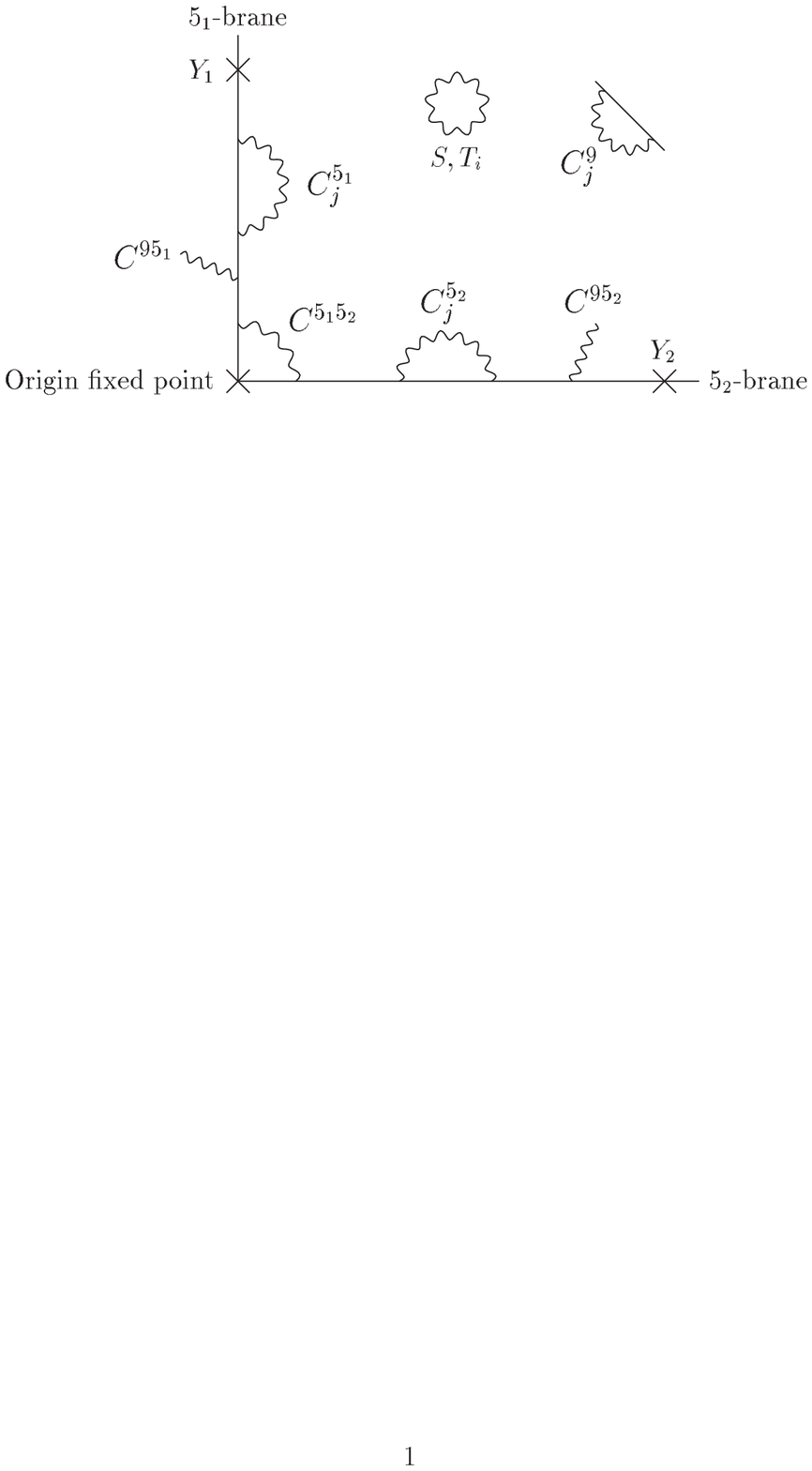}
\caption{\small A generic type I string construction involving two sets of
perpendicular D5-branes embedded within a D9-brane, where the D5-brane 
world-volumes intersect at the origin.  
Charged chiral fields appear as open strings with both ends attached to the 
same D-brane $C^{5_{i}}_{j}$ and $C^{9}_{j}$, or different branes 
$C^{5_{1} 5_{2}}$ and $C^{9 5_{i}}$.  Closed
strings ($S,T_{i}$) can live in the full 10d space, although orbifolding 
leads to closed strings (twisted moduli $Y_{k}$) localised at 4d fixed points 
within the $D5_{i}$-brane world-volume.}
\label{const}
\end{figure}

In order to study the effects of LFV elucidated in the previous 
section, it is necessary to specialize to a particular
effective non-minimal SUGRA model which addresses the 
question of flavour (i.e. provides a theory of the Yukawa couplings).
The specific model we shall discuss is 
defined in Table~\ref{tab:particle_content_42241}.
This model is an extension of the Supersymmetric
Pati-Salam model discussed 
in ref.\cite{Everett:2002pm}, based on two $D5$ branes which intersect
at $90$ degrees and preserve SUSY down to the TeV energy scale.
The generic D-brane set-up that we use is illustrated
in Fig.\ref{const}, where the string assignment notation
is defined. The gauge group of the $5_{1}$ sector is 
$U(4)^{(1)}\times U(2)^{(1)}_{L}\times U(2)^{(1)}_{R} $, 
and the gauge group of the
$5_{2}$  sector is \( U(4)^{(2)} \) (e.g., we assume the $U(2)_{L,R}$ of the
$5_2$ sector are broken).
The symmetry breaking pattern of this model takes place in two stages,
which we assume occur at very similar scales $\sim M_X$. In the first
stage, the $U(4)$ groups are broken to the diagonal subgroup via diagonal
VEV's of bifundamentals; the resulting theory is an effective   
Pati-Salam model (with additional $U(1)$'s) which then breaks to the MSSM
(and a number of additional $U(1)'s$) via the usual Higgs pair of
bifundamentals. The string scale is taken to be equal to the GUT scale, about
$3\times 10^{16}$ GeV. 

The symmetry breaking pattern leads to the following relations
among the gauge couplings of the SM gauge groups in terms of the gauge
couplings $g_{5_1}$ and $g_{5_2}$ associated with the gauge groups of the
$5_1$ and $5_2$ sectors: 
\begin{eqnarray}
\label{gaugecouplings3}
g_{3} & = & 
\frac{g_{5_{1}}g_{5_{2}}}{\sqrt{g^{2}_{5_{1}}+g^{2}_{5_{2}}}}=g_4\\
g_{2} & = & g_{5_{1}} =g_{2R}\\
g_{1} & = & \frac{\sqrt{3}g_{3}g_{2}}{\sqrt{3g^{2}_{3}+2g^{2}_{2}}}.
\label{gaugecouplings1}
\end{eqnarray}

\begin{table}[htbp]
  \centering \mbox{
  \begin{tabular}{|c||c|c|c|c||c|c|}
    \hline Field & $\mathrm{U}(4)^{(1)}$ & $\mathrm{U}(2)^{(1)}_L$ &
    $\mathrm{U}(2)^{(1)}_R$ & $\mathrm{U}(4)^{(2)}$ & Ends & $\UI_F$
    charge \\ 
    \hline $h$ & 1 & 2 & 2 & 1
    & $C_1^{5_1}$ & $0$ \\ 
    $F_3$ & 4 & 2 & 1 & 1 & $C_2^{5_1}$ &  $q_{L3}$  \\ 
    $\overline{F}_3$ & $\overline{4}$ & 1 & 2 &
    1 & $C_3^{5_1}$ & $q_{R3}$ \\ 
    $F_2$ & 1 & 2     & 1 & 4 & $C^{5_1 5_2}$ & $q_{L2}$ \\ 
    $\overline{F}_2$ & 1 & 1 & 2 & $\overline{4}$ & $C^{5_1 5_2}$ & $q_{R2}$  \\ 
    $F_1$ & 1 & 2 & 1 & 4 & $C^{5_1 5_2}$ & $q_{L1}$  \\
    $\overline{F}_1$ & 1 & 1 & 2 & $\overline{4}$ & $C^{5_1 5_2}$ &     $q_{R1}$ \\ 
    $H$ & 4 & 1 & 2 & 1 & $C_1^{5_1}$ &     $q_H$ \\ 
    $\overline{H}$ & $\overline{4}$ & 1 & 2 & 1 & $C_2^{5_1}$ & $\!\!\!\!\!-q_H$\\ 
    $\varphi_1$ & 4 & 1 & 1 & $\overline{4}$ & $C^{5_1 5_2}$ & $-$ \\
    $\varphi_2$ & $\overline{4}$ & 1 & 1 & 4 & $C^{5_1 5_2}$ & $-$ \\ 
    $D_6^{(+)}$ & 6 & 1 & 1 & 1 & $C_1^{5_1}$ & $-$  \\
    $D_6^{(-)}$ & 6 & 1 & 1 & 1 & $C_2^{5_2}$ & $-$  \\ 
    $\theta$  & 1 & 1 & 1 & 1 & $C^{5_1 5_2}$ & $\!\!\!\!\!-1$ \\
    $\overline{\theta}$ & 1 & 1 & 1 & 1 & All & $1$ \\
    \hline
  \end{tabular} }
  \caption{{\small The particle content of the 42241 model, and the brane
assignments of the corresponding string. Note that the string
assignment of $\overline{\theta}$ is allowed to be any of $\{ C^{5_1
5_2}, C^{5_1}_1, C^{5_1}_2, C^{5_1}_3\}$, giving a slightly different
model in each case.  The $U(1)_F$ charges will change, also generating
a slightly different model in each case. We will return to this in
more detail in Section~\ref{sec:numerical-etc}. } }
  \label{tab:particle_content_42241}
\end{table}

The extension is to include an additional $U(1)_F$ family symmetry and
the FN operators responsible for the Yukawa couplings
as in \cite{Blazek:2003wz} (see also
\cite{King:OperatorAnalysis}).  
The charges under the Abelian symmetry $U(1)_F$ are left arbritary for
now. The present `42241' Model is the same
as the model considered in \cite{KP0307091}, with the following
modifications considered; firstly, we allow the Froggatt-Nielsen field
$\overline{\theta}$ to be either an intersection state or attached to
the $5_1$ brane. The location of $\overline{\theta}$ dramatically
changes the value of the D-term contribution to the scalar masses
coming from the FN sector.

The quark and lepton fields are contained in the representations 
$F, \overline{F}$ which are assigned
charges under $\UI_F$. In Table~\ref{tab:particle_content_42241}
we list the charges, string assignments and representations under the
string gauge group 
$U(4)^{(1)} \otimes U(2)^{(1)}_L \otimes U(2)^{(1)}_R \otimes U(4)^{(2)}$.

The field $h$ represents both Electroweak Higgs doublets that
we are familiar with from the MSSM.  The fields $H$ and $\overline{H}$
are the Pati-Salam Higgs scalars;\footnote{We will also refer to these as
``Heavy Higgs''; this has nothing to do with the MSSM heavy neutral higgs
state $H^0$.} the bar on the second is used to note that it is in the conjugate
representation compared to the unbarred field.

The extra Abelian $U(1)_F$ gauge group is a family symmetry, and is broken at
the high energy scale by the \vevs{} of the FN fields \cite{Froggatt:1978nt} 
$\theta,\overline{\theta}$, which have charges $-1$ and $+1$ respectively
under $U(1)_F$. 
We assume that the singlet field $\theta$ 
arises as an intersection
state between the two \Dbranes{5}, transforming under the remnant
\UI s in the 4224 gauge structure. In general the FN fields are expected to 
have non-zero F-term \vev s.

The two $SU(4)$ gauge groups are broken to their diagonal
subgroup at a high scale due to the assumed \vevs{} of the
bifundamental
Higgs fields $\varphi_1$, $\varphi_2$ \cite{Everett:2002pm}.
The symmetry breaking at the scale $M_X$
\begin{equation}
  \label{eq:gauge_breaking_pattern}
  \psgroup \rightarrow \smgroup
\end{equation}
is achieved by the heavy Higgs fields $H$, $\overline{H}$
which are assumed to gain \vevs{} \cite{King:OperatorAnalysis}
\begin{equation}
  \label{eq:heavy_higgs_\vevs}
  \left< H^{\alpha b} \right> = \left< \nu_H \right> = V
  \delta^\alpha_4 \delta^b_2 \sim M_X \;\;\; ;\;\;\; \left<
  \overline{H}_{\alpha x} \right> = \left< \overline{\nu}_H \right> =
  \overline{V} \delta_\alpha^4 \delta_x^2 \sim M_X
\end{equation}
This symmetry breaking splits the Higgs field $h$
into two Higgs doublets, $h_1$, $h_2$. Their neutral components then gain
weak-scale \vevs{}.
\begin{equation}
  \label{eq:mssm_higgs_\vevs}
  \left< h_1^0 \right> = v_1 \;\;\; ; \;\;\;
  \left< h_2^0 \right> = v_2 \;\;\; ; \;\;\;
  \tan \beta = v_2 / v_1.
\end{equation}
The low energy limit of this model contains the MSSM with right-handed
neutrinos.  We will return to the right handed neutrinos when we
consider operators including the heavy Higgs fields $H$,
$\overline{H}$ which lead to effective Yukawa contributions and
effective Majorana mass matrices when the heavy Higgs fields gain
\vevs.


\subsection{Yukawa operators}
\label{sec:yukawa-sector}


The (effective) Yukawa couplings are generated by operators
involving the FN field $\theta$ with
the following structure:\footnote{The field $\overline{\theta}$ will not
enter the Yukawa operators because $F_i \overline{F}_j h$ will be positive for
any $i,j$.}
\cite{King:OperatorAnalysis}:
\begin{equation}
  \label{eq:dirac_operator_n_is_n}
  \mathcal{O} = F_i \overline{F}_j h \left(\frac{H
  \overline{H}}{M_X^2}\right)^n
  \left(\frac{\theta}{M_X}\right)^{p(i,j)}
\end{equation}
where the integer
$p(i,j)$ is the total $U(1)_F$ charge of $F_i + \overline{F}_j + h$ and
$H\overline{H}$ has a $U(1)_F$ charge of zero.
The tensor structure of the operators in
Eq.~(\ref{eq:dirac_operator_n_is_n}) is
\begin{equation}
  \label{eq:dirac_mass_tensor}
  \left(\mathcal{O}\right)^{\alpha\rho y w}_{\beta\gamma x z} =
  F^{\alpha a} \overline{F}_{\beta x} h^y_a \overline{H}_{\gamma z}
  H^{\rho w} \theta^{p(i,j)}
\end{equation}
One constructs
$\SU{4}_{PS}$ invariant tensors $C^{\beta\gamma}_{\alpha\rho}$ that
combine $4$ and $\overline{4}$ representations of $\SU{4}_{PS}$ into
\boldmath $1$, $6$, $10$, $\overline{10}$ and $15$ \unboldmath
representations \cite{King:OperatorAnalysis}. Similarly we construct $\SU{2}_R$
tensors $R^{xz}_{yw}$ that combine the $\mathbf{2}$ representations of \SU{2} into
singlet and triplet representations. These tensors are contracted
together and into $\mathcal{O}^{\alpha\rho y w}_{\beta \gamma
x z}$ to create singlets of $\SU{4}_{PS}$, $\SU{2}_L$ and $\SU{2}_R$.
Depending on which operators are used, different
Clebsch-Gordan coefficients (CGCs) will emerge.

We will return to these in section \ref{sec:numerical-etc}, when we define
the two models that we will be using for the numerical analysis.


\subsection{Majorana operators}
\label{sec:majorana-fermions}


We are interested in Majorana fermions because they can contribute
neutrino masses of the correct order of magnitude via the see-saw
effect. The operators for Majorana fermions are of the form

\begin{equation}
  \label{eq:majorana_operators_n_is_n}
  \mathcal{O}_{ij} = \overline{F}_i \overline{F}_j \left(\frac{H H}{M_X}\right)
  \left(\frac{H \overline{H}}{M_X^2}\right)^{n-1}
  \left(\frac{\theta}{M_X}\right)^{q(i,j)}
\end{equation}

There do not exist renormalisable elements of this infinite series of
operators, so $n < 1$ Majorana operators are not defined, except in
the $(3,3)$ element. We assume that a $(3,3)$ neturino mass term is allowed at leading
(but non-renormalisable) order.
A similar analysis goes through
as for the Dirac fermions; however the structures only ever give
masses to the neutrinos, not to the electrons or to the
quarks.


\section{Soft supersymmetry breaking masses}
\label{soft}


\subsection{Supersymmetry breaking F-terms}


In \cite{Everett:2002pm} it was assumed that the Yukawas were
field-independent, and hence the only $F$-vevs of importance were that
of the dilaton ($S$), and the untwisted moduli ($T^i$).
Here we set out the parameterisation for the F-term \vevs{},
including the contributions from the FN field $\theta$ and the
heavy Higgs fields $H,\overline{H}$. Note that
the field dependent part follows from the assumption that the
family symmetry field, $\theta$ is an intersection state.
\begin{eqnarray}
  \label{eq:dilaton_auxilliary_vev_42241}
  F_S
  &=&
  \sqrt{3} m_{3/2} \left( S + \overline{S} \right)
  X_S
  \\
  \label{eq:untwisted_mod_auxilliary_vev_42241}
  F_{T_i}
  &=&
  \sqrt{3} m_{3/2} \left( T_i + \overline{T}_i \right)
  X_{T_i}
  \\
  \label{eq:heavy_higgs_auxilliary_vev_42241}
  F_{H^{\alpha b}} 
  &=&
  \sqrt{3} m_{3/2} H^{\alpha b} \left( S + \overline{S} \right)^\frac{1}{2}
  X_H
  \\
  \label{eq:heavy_conj_higgs_auxilliary_vev_42241}
  F_{\overline{H}_{\alpha x}} &=& \sqrt{3} m_{3/2}
  \overline{H}_{\alpha x} \left( T_3 + \overline{T}_3
  \right)^\frac{1}{2} X_{\overline{H}} \\
  \label{eq:family_field_auxilliary_vev_42241}
  F_\theta
  &=&
  \sqrt{3} m_{3/2} \theta \left(S + \overline{S}\right)^\frac{1}{4}
  \left(T_3 + \overline{T}_3\right)^\frac{1}{4}
  X_\theta
\end{eqnarray}
We introduce a shorthand notation:
\begin{equation}
  \label{eq:shorthand_key}
  F_H H = \sum_{\alpha b} F_{H^{\alpha b} } H^{\alpha b} \; ;\;
  F_{\overline{H}} \overline{H} = \sum_{\alpha x}
  F_{\overline{H}_{\alpha x} } \overline{H}_{\alpha x}.
\end{equation}

The F-terms above use values of $S$ and $T_i$ which 
are given in terms of the gauge couplings as:
\begin{equation}
\mathrm{Re}(S) = \frac{4\pi}{g^2_9},\ \ 
\mathrm{Re}(T_i) = \frac{4\pi}{g^2_{5_i}}.
\end{equation}
The gauge couplings $g_{5_1},g_{5_2}$ are given from 
Eq.~(\ref{gaugecouplings3})-(\ref{gaugecouplings1}) \cite{Everett:2002pm} as
\begin{equation}
g_{5_1} = g_2, \ \ g_{5_2} = \frac{g_2 g_3}{\sqrt{g^2_2
    -g^2_3}},
\end{equation} 
where we shall assume that at the scale $M_X$ we have,
\begin{equation}
g_2 = 0.7345, \ \ g_3 = 0.6730 .
\end{equation}
The values of $g_{9},g_{5_3}$ are assumed to be equal and
are obtained from the string relation
\begin{equation}
\label{eq:string gauge couplings}
32 \pi^2 \left( \frac{M_*}{M_{Pl}} \right)^2 = g_9 g_{5_1} g_{5_2} g_{5_3} ,
\end{equation}
as 
\begin{equation}
g_9 = g_{5_3} = 0.0266 ,
\end{equation}
where we have taken
\begin{equation}
\left( \frac{M_*}{M_{Pl}} \right)^2 = 2.77\times10^{-6}.
\end{equation}
These rather small gauge couplings imply
\begin{equation}
\mathrm{Re}(S) = \mathrm{Re}(T_3) = 0.877.
\end{equation}
In \cite{KP0307091} the string relation was not used and 
it was assumed incorrectly that 
$g_9 = g_{5_3} = g_2 $ which resulted in 
$\mathrm{Re}(S) = 27.7$.


\subsection{Soft scalar masses}
\label{sec:scalars_42241}


There are two contributions to scalar mass
squared matrices, coming from SUGRA and from D-terms.
In this subsection we calculate the SUGRA predictions for the
matrices at the GUT scale, and in the next subsection we add on the 
D-term contributions.

The SUGRA contributions to soft masses are detailed in 
Section~\ref{sec:soft-from-SUGRA}.

From Eq.~(\ref{eq:normalised_scalars})  we can get the family independent
form for all scalars:
\begin{eqnarray}
  \label{eq:scalars_left}
  m^2_{L} &=& m^2_{3/2}
  \left[
    \begin{array}{ccc}
      a & & \\
      & a & \\
      & & b_L
    \end{array}
  \right] \\ \label{eq:scalars_right}
  m^2_{R} &=& m^2_{3/2}
  \left[
    \begin{array}{ccc}
      a & & \\
      & a & \\
      & & b_R
    \end{array}
  \right] \\
  \label{eq:mssm_higgs_at_mx}
  m^2_h &=& m^2_{3/2} ( 1 - 3 X^2_S )\\
  \label{eq:ps_H_at_mx}
  m^2_H &=& m^2_{3/2} ( 1 - 3 X^2_S ) \\
  \label{eq:ps_Hbar_at_mx}
  m^2_{\overline{H}} &=& m^2_{3/2}(1 - 3 X^2_{T_3} ) \\
  \label{eq:FN soft mass}
  m^2_\theta &=& m^2_{3/2} \left[ 1 - \frac{3}{2} (X^2_S + X^2_{T_3} ) \right]
\end{eqnarray}
where
\begin{eqnarray}
  \label{eq:SUGRA_a}
  a &=& 1 - \frac{3}{2} (X^2_S + X^2_{T_3} ) \\
  \label{eq:SUGRA_b_L}
  b_L &=& 1 - 3 X^2_{T_3} \\
  \label{eq:SUGRA_b_R}
  b_R &=& 1 - 3 X^2_{T_2}
\end{eqnarray}
Here $m^2_L$ represents the left handed scalar mass squared
matricies $m^2_{Q_L}$ and $m^2_{L_L}$. $m^2_R$ represents the right
handed scalar mass squared matricies $m^2_{U_R}$, $m^2_{D_R}$,
$m^2_{E_R}$ and $m^2_{N_R}$. A discussion of the equations for
$m^2_{\overline{\theta}}$ can be found in Section~\ref{sec:theta_bar D-terms}.


\subsection{D-term contributions}
\label{sec:D-term_contributions}


There are two D-term contributions to the scalar masses. The first is the well
known \cite{King:2000vp,KP0307091} contribution from the breaking of the Pati-Salam
group to the MSSM group. Note that these D-terms are different to those quoted
in the references above as we now consider the D-terms generated by breaking a
family symmetry. See Appendix~\ref{sec:deriv D-terms} for a full derivation.
The second D-term comes solely from the breaking of the $U(1)$ family
symmetry. The corrections lead to the following mass matrices:

\begin{eqnarray}
\label{eq:m^2_QL}
m^2_{Q_L} &=& m^2_L + {\bf 1} (g^2_4) D^2_H +
\begin{pmatrix}
q_{L1} & & \\ & q_{L2} & \\ & & q_{L3}
\end{pmatrix}
g^2_F D^2_\theta \; , \\
\label{eq:m^2_LL}
m^2_{L_L} &=& m^2_L -{\bf 1} (3g^2_4) D^2_H +
\begin{pmatrix}
q_{L1} & & \\ & q_{L2} & \\ & & q_{L3}
\end{pmatrix}
g^2_F D^2_\theta \; , \\
\label{eq:m^2_UR}
m^2_{U_R} &=&  m^2_R -{\bf 1}(g^2_4 -2g^2_{2R}) D^2_H +
\begin{pmatrix}
q_{R1} & & \\ & q_{R2} & \\ & & q_{R3}
\end{pmatrix}
g^2_F D^2_\theta \; , \\
\label{eq:m^2_DR}
m^2_{D_R} &=&  m^2_R -{\bf 1} (g^2_4 +2g^2_{2R}) D^2_H +
\begin{pmatrix}
q_{R1} & & \\ & q_{R2} & \\ & & q_{R3}
\end{pmatrix}
g^2_F D^2_\theta \; , \\
\label{eq:m^2_ER}
m^2_{E_R} &=&  m^2_R +{\bf 1} (3g^2_4 -2g^2_{2R}) D^2_H +
\begin{pmatrix}
q_{R1} & & \\ & q_{R2} & \\ & & q_{R3}
\end{pmatrix}
g^2_F D^2_\theta \; , \\
\label{eq:m^2_NR}
m^2_{N_R} &=&  m^2_R +{\bf 1} (3g^2_4 +2g^2_{2R}) D^2_H +
\begin{pmatrix}
q_{R1} & & \\ & q_{R2} & \\ & & q_{R3}
\end{pmatrix}
g^2_F D^2_\theta \; , \\
\label{eq:m^2_hu}
m^2_{h_u} &=& m^2_{h_2} - 2g^2_{2R} D^2_H \; , \\
\label{eq:m^2_hd}
m^2_{h_d} &=& m^2_{h_1} + 2g^2_{2R} D^2_H \; ,
\end{eqnarray}

The charges $q_{Li}, q_{Rj}$ are the charges under $U(1)_F$ of $F_i$
and $\overline{F}_j$ respectively, as shown in
Table~\ref{tab:charges_both_models}.  The correction factors
$D^2_\theta, D^2_H$ are calculated explicitly in Appendix
\ref{sec:deriv D-terms} in terms of the gauge couplings and soft
masses as\footnote{ $q_H$ is defined to be $-q_{R3}$, thus $q_H = \frac{5}{6}$
for Model 1 and $q_H = 1$ for Model 2. See Appendix~\ref{sec:deriv D-terms} for
more details.}
\begin{eqnarray}
\label{eq:1}
D^2_H &=& \frac{1}{4 g^2_{2R} + 6 g^2_4} \left[ m^2_H -
m^2_{\overline{H}} + q_H(m^2_\theta - m^2_{\overline{\theta}})\right] \\
\label{eq:2}
D^2_\theta &=& \frac{m^2_\theta - m^2_{\overline{\theta}}}{2 g_F^2}
\end{eqnarray}

We note that the factors of $g^2_F$ appearing in the mass matrices are
cancelled by the $\frac{1}{g^2_F}$ in the definition of $D^2_\theta$.

The D-terms associated with the family symmetry depend on the charges
of the left-handed and right-handed matter representations
$F,\overline{F}$ under the family symmetry. It is well
known\footnote{For an explanation, see for example \cite{Kane:2005va}.} that
for Pati-Salam, one can choose any set of charges, and there will be
an equivalent, shifted set of charges that are anomaly free due to
the Green-Schwartz anomaly cancellation mechanism. The charges used for
the D-term calculation should be the anomaly free charges.

The gauge couplings and mass parameters in
Eqs.~(\ref{eq:1}~,~\ref{eq:2}) are predicted from the model, in terms
of the $X$ parameters and $m_{3/2}$ as shown in
Eqs.~(\ref{eq:ps_H_at_mx}~--~\ref{eq:FN soft mass}) and
Eqs.~(\ref{eq:C^5_1 _1 D^2_theta}~--~\ref{eq:C^5_1 _3 D^2_theta}).
Note that the D-terms will be zero if $X_S = X_{T_i}$, or if the
$\overline\theta$ brane assignment is the same as $\theta$. Choosing
the second of these conditions is useful since it gives a comparison
case where there are no $U(1)_F$ D-terms; this comparison will make
the D-term contribution to flavour violation immediately apparent.


\subsection{Magnitude of $D_{\theta}$-terms for different
$\overline{\theta}$ assignments}
\label{sec:theta_bar D-terms}


The main point worth emphasising is that in the string model
the magnitudes of the $D$-terms are {\em calculable}.
We have assumed throughout that the FN field $\theta$ is an
intersection string state $C^{5_15_2}$, but have not specified the
string assignment of $\overline{\theta}$.
Thus $m^2_{\overline{\theta}}$ takes various values depending on the
string assignment for $\overline{\theta}$.

From Eq.~(\ref{eq:2}), we see that we have calculable D-terms,
\begin{equation}
\label{eq:D^2_theta goes like}
2g^2_F D^2_\theta = m^2_\theta -m^2_{\overline{\theta}} \; ,
\end{equation}
so the value of $D^2_\theta$ depends on the choice of where the $\overline{\theta}$
field lives. We use Table~\ref{tab:particle_content_42241} and
Eqs.~(\ref{eq:scalars_left} - \ref{eq:SUGRA_b_R}) to quantify the
$D_\theta$-term for each possible $\overline{\theta}$ string assignment.
As $\theta$ always lives at the intersection, on $C^{5_1 5_2}$, our first choice of
$\overline{\theta}$ on $C^{5_1 5_2}$ is trivial: it gives $D^2_\theta =0$.
For $\overline{\theta}$ on $C^{5_1}_1$, $m^2_{\overline{\theta}}$ is equivalent to
$m^2_h$, as this is also on $C^{5_1}_1$. So using Eq.~(\ref{eq:mssm_higgs_at_mx})
for $m^2_{\overline{\theta}}$ and Eq.~(\ref{eq:FN soft mass}) for $m^2_\theta$
in Eq.~(\ref{eq:D^2_theta goes like}), we have
\begin{eqnarray}
\label{eq:C^5_1 _1 D^2_theta}
C^{5_1}_1 &:& 2g^2_F D^2_\theta = \frac{3}{2} m^2_{3/2} (X^2_S - X^2_{T_3}) .
\end{eqnarray}
Similarly, the other two choices yield
\begin{eqnarray}
\label{eq:C^5_1 _2 D^2_theta}
C^{5_1}_2 &:& 2g^2_F D^2_\theta = -\frac{3}{2} m^2_{3/2} (X^2_S - X^2_{T_3}) \\
\label{eq:C^5_1 _3 D^2_theta}
C^{5_1}_3 &:& 2g^2_F D^2_\theta = -\frac{3}{2} m^2_{3/2} (X^2_S - X^2_{T_2}) .
\end{eqnarray}

In this paper, all models use $X^2_{T_3} = X^2_{T_2} = X^2_{T_1} = X^2_T$, so
Eqs.~(\ref{eq:C^5_1 _2 D^2_theta}) and (\ref{eq:C^5_1 _3 D^2_theta}) are
equal to each other, and opposite in sign to Eq.~(\ref{eq:C^5_1 _1 D^2_theta}).


\subsection{Soft gaugino masses}
\label{sec:param-gaug}


The soft gaugino masses are the same as in
\cite{Everett:2002pm}, which we quote here for completeness.
The results follow from
Eq.~(\ref{eq:normalised_gauginos}) applied to the \psgroup gauginos, which
then mix into the \smgroup gauginos whose masses are given by
\begin{eqnarray}
  \label{eq:su3_gaugino_gut_scale}
  M_3 &=& \frac{\sqrt{3} m_{3/2} } {\left(T_1 + \overline{T}_1\right)
  + \left(T_2 + \overline{T}_2\right) } \left[ (T_1 + \overline{T}_1)
  X_{T_1} + (T_2 + \overline{T}_2 ) X_{T_2} \right] \\
 \label{eq:su2_gaugino_gut_scale}
  M_2 &=& \sqrt{3} m_{3/2} X_{T_1} \\
  \label{eq:u1_gaugino_gut_scale}
  M_1 &=& \frac{\sqrt{3} m_{3/2} } {\frac{5}{3} (T_1 + \overline{T}_1
  ) + \frac{2}{3} (T_2 + \overline{T}_2 ) } \left[ \frac{5}{3}(T_1 +
  \overline{T}_1) X_{T_1} + \frac{2}{3}(T_2 + \overline{T}_2 ) X_{T_2}
  \right]
\end{eqnarray}
The values of $T_1 + \overline{T}_1$ and
$T_2 + \overline{T}_2$ are proportional to the brane
gauge couplings $g_{5_1}$ and $g_{5_2}$, which are related in a simple
way to the MSSM couplings at the unification scale. This is discussed
in \cite{Everett:2002pm}.

When we run the MSSM gauge couplings up and solve for $g_{5_1}$ and
$g_{5_2}$ we find that approximate gauge coupling unification is 
achieved by $T_1 +\overline{T_1} \gg T_2 +
\overline{T}_2$. Then we find the simple approximate result
\begin{equation}
  \label{eq:63}
  M_1 \approx M_3 \approx M_2 = \sqrt{3}m_{3/2} X_{T_1}.
\end{equation}


\subsection{Soft trilinear couplings}
\label{sec:param_tril}


So far we have considered the soft masses and the gaugino masses. The gaugino
masses are the same as in \cite{Everett:2002pm}. The soft masses have had
both D-term contributions added onto the base values from \cite{Everett:2002pm}.
The contributions to the soft masses and gaugino masses from the FN and heavy Higgs
auxiliary fields is completely negligible due to the small size of their F-terms.
However for the soft trilinear masses 
these contributions are of order $\mathcal{O}(m_{3/2})$ despite 
having small F-terms, so FN and Higgs contributions will
give very important additional contributions beyond those
considered in \cite{Everett:2002pm}.

From Section~\ref{sec:soft-from-SUGRA} we see that
the canonically normalised equation for the trilinear is
\begin{equation}
  \label{eq:normalised_trilinear0}
  A_{abc} = F_I 
  \left[
    \overline{K}_I - \partial_I \ln
    \left(
      \tilde K_a \tilde K_b \tilde K_c
    \right)
  \right]
  + F_m \partial_m \ln Y_{abc}
\end{equation}
This general form for the
trilinear accounts for contributions from non-moduli F-terms. These
contributions are in general expected to be of the same magnitude as
the moduli contributions despite the fact that the non-moduli F-terms
are much smaller \cite{Abel:2001cv}. Specifically,
if the Yukawa hierarchy is taken to be generated by a FN
field, $\theta$ such that $Y_{ij} \sim \theta^{p_{ij}}$, 
then we expect $F_\theta \sim m_{3/2} \theta$, 
and then $\Delta A_{ij}=F_\theta \partial_\theta \ln Y_{ij} \sim
p_{ij}m_{3/2}$ and so even though these fields are expected to have heavily
sub-dominant F-terms\footnote{In our model the FN and heavy Higgs vevs are 
of order the unification scale, compared to the moduli vevs
which are of order of the Planck scale.}
they contribute to the trilinears at the same
order $\mathcal{O}(m_{3/2})$ as the moduli, but in a flavour off-diagonal way.

Here we sum over $m$,
which contains all the hidden sector fields: $S, T_i, H, \overline{H}, \theta$.

In the specific D-brane model of interest here, the general results for
soft trilinear masses, including the contributions for general effective Yukawa
couplings are given in Appendix \ref{sec:param-tril-42241}. 
From Eqs.~(\ref{eq:dirac_operator_n_is_n}, \ref{eq:dirac_mass_tensor})
we can read off the effective Yukawa couplings,
\begin{equation}
  \label{eq:dirac_yukawa}
  Y_{h F \overline{F} } h F \overline{F} \equiv
  \underbrace{(c)^{\beta\gamma}_{\alpha\rho}(r)^{xz}_{yw}\overline{H}_{\gamma
  z} H^{\rho w}\theta^{p}}_{ {Y_{hF\overline{F}}}^{\beta
  x}_{\alpha y}}h^y_a F^{\alpha a} \overline{F}_{\beta x}.
\end{equation}
Note the extra group indices that the effective Yukawa coupling
${Y_{hF\overline{F}}}^{\beta x}_{\alpha y}$ has, and 
proper care must be taken of the tensor structure
when deriving trilinears from a given operator.
We can write down the
trilinear soft masses, $A$, by substituting the operators in
Eq.~(\ref{eq:operator_texture1}) into the results in 
Appendix \ref{sec:param-tril-42241}.
Having done this we find the result:\footnote{We assume that a $(3,3)$ Yukawa
coupling appears at renormalisable order. This is why the $A_{33}$ doesn't include
contributions from $d_H$ and $d_\theta$.} 
\begin{equation}
  \label{eq:trilinear}
   A = \sqrt{3} m_{3/2}
   \left[
    \begin{array}{ccc}
      d_1 + d_H + p(i,j)d_\theta & d_1 + d_H + p(i,j)d_\theta
      & d_2 + d_H + p(i,j)d_\theta \\
      d_1 + d_H + p(i,j)d_\theta & d_1 + d_H + p(i,j)d_\theta
      & d_2 + d_H + p(i,j)d_\theta \\
      d_3 + d_H + p(i,j)d_\theta & d_3 + d_H + p(i,j)d_\theta & d_4
    \end{array}
  \right]  
\end{equation}
where
\begin{eqnarray}
    d_1 &=& X_S - X_{T_1} - X_{T_2} \\ d_2 &=& \frac{1}{2} X_S -
  X_{T_1} - \frac{1}{2}X_{T_2} \\ d_3 &=& \frac{1}{2} X_S - X_{T_1} -
  X_{T_2} + \frac{1}{2}X_{T_3} \\ d_4 &=& -X_{T_1} \\ d_H &=&
  (S+\overline{S})^\frac{1}{2} X_H + (T_3 +
  \overline{T}_3)^\frac{1}{2} X_{\overline{H}} \\ d_\theta &=&
  (S+\overline{S})^\frac{1}{4} (T_3 + \overline{T}_3)^\frac{1}{4}
  X_\theta
\end{eqnarray}

These results are independent of which string assignment we give to
$\overline\theta$, since this field does not enter the Yukawa operators.


\section{Results}
\label{sec:numerical-etc}


\subsection{Two models}
\label{sec:models}

We will study two different models which have the same Yukawa
textures, but different $U(1)_F$ charge structures. 
The anomaly-free \cite{Kane:2005va}
family charges are laid out in Table~\ref{tab:charges_both_models}.
This first model is essentially the model studied in
\cite{Blazek:2003wz,KP0307091}, but with an extra operator in the
$(1,2)$ and $(1,3)$ Yukawa matrix element to allow a non-zero $Y^e_{12}$ 
and $Y^e_{13}$. 
The second model is defined such that all of the
charges of left-handed matter are the same, causing the $U(1)_F$
D-term to left handed scalar mass matrices to not lead to extra
flavour violation. This choice is a small change to the model, since
two of the `left-handed' charges are the same anyway, and is made
because normally the left-handed contribution dominates over the
right-handed contribution. This can be
achieved by changing the order of the operators in the Yukawa textures
and the arbritary couplings $a,a', ...$ to compensate the change in
the charge structure, as discussed in Appendix~\ref{operatorsformodels}.
Most of the results presented will be for Model 1. 

Model 2 differs from Model 1 in the charges under the Abelian
family symmetry, and the compensating changes to ensure the same
Yukawa textues. As discussed in Appendix~\ref{operatorsformodels}
there are two expansion parameters in these models which 
we take to be equal $\epsilon = \delta = 0.22$. The powers
of $\epsilon$ in the first row of the Yukawa matrices
are one lower, and we compensate that by
increasing the powers of $\delta$ in the first row. We do not have to
shift the values of the $a,a',a'',a'''$ parameters since $\delta =
\epsilon$. Were this not the case, we would have to shift the values
by a factor of $\frac{\epsilon}{\delta}$.  Model 2 is not meant to be
a natural or realistic model, we use it as a tool to investigate the
contribution to flavour violation from the $U(1)_F$ D-term correction
to the left-handed scalar masses. The family charges are laid out in
Table~\ref{tab:charges_both_models}.

The $U(1)_F$ charge of $H$ must be equal and opposite to
$\overline{F}_3$, and $\overline{H}$ must be the negative of
this. This is due to the (3,3) element of the right-handed Majorana
mass being allowed at leading order, so the $U(1)_F$ charges of
$\overline{F}_3$ and $H$ must conspire to cancel for the operator of
the Majorana fermions to be renormalisable.
For Model 1, the $U(1)_F$ charges of $H$ and $\overline{H}$ are $5/6$
and $-5/6$ respectively. 
For Model 2, the $U(1)_F$ charges of $H$ and $\overline{H}$ are $1$
and $-1$ respectively. One can use the relevent equations above to
check that the anomaly coefficients do indeed satisfy the anomaly
cancellation conditions.
This gives us a different $D_H$-term for Model 2, as indicated by $q_H$ in
Eq.~(\ref{eq:1}) being different for Model 1 and Model 2. Note that
$D^2_\theta$ is the same in both models.

\begin{table}[htbp]
  \centering
  \begin{tabular}{|c|c|c|}
    \hline
    Field  & Model 1 & Model 2 \\ 
    & Charge & Charge  \\
    \hline
    $F_1$ & $\frac{11}{6}$ & $1$ \\
    $F_2$ & $\frac{5}{6}$ & $1$ \\
    $F_3$ & $\frac{5}{6}$ & $1$ \\
    \hline
    $\overline{F}_1$ & $\frac{19}{6}$ & $3$ \\
    $\overline{F}_2$ & $\frac{7}{6}$ & $1$ \\
    $\overline{F}_3$ & $\!\!\!\!-\frac{5}{6}$ & $\!\!\!\!-1$ \\
    \hline
  \end{tabular}
  \caption{The family charges for Model 1 and Model 2.}
  \label{tab:charges_both_models}
\end{table}

The values of the arbritary couplings are laid out in 
Appendix~\ref{operatorsformodels} in
Table~\ref{tab:a_ap_app_for_models}. This gives numerical values for
the Yukawa elements which can be used in either model, with the relevant
values of $Y^e_{12}$ and $Y^e_{13}$ inserted instead of the texture zeros:
\begin{eqnarray}
  \label{eq:numerical_yu_model1}
  Y^{u}(M_X) &=&
  \left[
    \begin{array}{lll}
      2.159\times10^{-06}  &  5.606\times10^{-04}  &  5.090\times10^{-03} \\
      0.000    &    1.105\times10^{-03}  &   0.000    \\
      0.000    &    6.733\times10^{-03}  &  5.841\times10^{-01}    \\
    \end{array}
  \right]
  \\
  \label{eq:numerical_yd_model1}
  Y^{d}(M_X) &=&
  \left[
    \begin{array}{lll}
    \!\!\!\!-1.661\times10^{-04} & \!\!\!\! -5.606\times10^{-04} &   1.018\times10^{-02} \\
     7.683\times10^{-04}  & \!\!\!\!-5.343\times10^{-03}  &  1.216\times10^{-02}\\
    \!\!\!\!-1.769\times10^{-04}   & 3.133\times10^{-02}   & 3.933\times10^{-01}\\
    \end{array}
  \right]
  \\
  \label{eq:numerical_ye_model1}
  Y^{e}(M_X) &=&
  \left[
    \begin{array}{lll}
    \!\!\!\!-1.246\times10^{-04} &    0.000   &      0.000   \\
     1.537\times10^{-03}  &  2.432\times10^{-02}&   \!\!\!\!-3.649\times10^{-02}\\
    \!\!\!\!-1.327\times10^{-04}   & 3.133\times10^{-02} &   5.469\times10^{-01}\\
  \end{array}
\right]
  \\
  \label{eq:numerical_yn_model1}
  Y^{\nu}(M_X) &=&
  \left[
    \begin{array}{lll}
     2.159\times10^{-06}&    1.525\times10^{-03}&    0.000\\
      0.000     &   8.290\times10^{-04} &   3.923\times10^{-01}  \\  
      0.000      &  5.050\times10^{-03}  &  5.469\times10^{-01} \\
    \end{array}
  \right]
\end{eqnarray}

The RH Majorana neutrino mass matrix for Models 1 and 2 has the 
numerical values:
\begin{equation}
  \label{eq:numerical_21}
  \frac{M_{RR}(M_X)}{M_{33}} =
  \left[
    \begin{array}{lll}
      3.508\times10^{8}  &       3.686\times10^{9}   &     3.345\times10^{11}\\
      3.686\times10^{9}   &     8.313\times10^{10}&    5.886\times10^{12}\\
     3.345\times10^{11}&    5.886\times10^{12} &   5.795\times10^{14}\\
    \end{array}
  \right]
\end{equation}



\subsection{Benchmark points}


Since the parameter space for the models is reasonably expansive, and
the intention is to compare different sources of LFV, it is convenient
to consider five benchmark points, as follows.
It should be noted that for all these points, we have taken all $X_{T_i}$
to be the same, $X_{T_i} = X_T$, and also $X_H = X_{\overline{H}}$.
$X_{\overline{\theta}}$ is taken to be zero throughout.

\label{sec:four_points}
\begin{table}[htbp]
  \centering
  \begin{tabular}{|c|c|c|c|c|c|c|}
    \hline
    Point & $X_S$ & $X_T$ & $X_H$ & $X_{\overline{H}}$ & $X_\theta$ &
    $X_{\overline{\theta}}$ \\
    \hline
    A & 0.500 & 0.500 & 0.000 & 0.000 & 0.000 & 0.000 \\
    B & 0.535 & 0.488 & 0.000 & 0.000 & 0.000 & 0.000 \\
    C & 0.270 & 0.270 & 0.000 & 0.000 & 0.841 & 0.000 \\
    D & 0.270 & 0.270 & 0.595 & 0.595 & 0.000 & 0.000 \\
    \hline
    \hline
    E & 0.290 & 0.264 & 0.000 & 0.000 & 0.841 & 0.000 \\
    \hline
  \end{tabular}
  \caption{\small 
Values of the X parameters for the five benchmark points, A-E.}
  \label{tab:benchmarks_defined}
\end{table}

\begin{itemize}
\item
  Point A is referred to as ``minimum flavour violation''.  At the
  point $X_S=X_T$ the scalar mass matrices $m^2$ are proportional to the
  identity, and the trilinears $\tilde A$ are aligned with the
  Yukawas. Also, if we look back to Eqs.~(\ref{eq:1},\ref{eq:2}),
  (\ref{eq:ps_H_at_mx}) and (\ref{eq:ps_Hbar_at_mx}), for
  $X_S = X_T$, which is the case for point A (and points C and D) we see
  that the value of both D-term contributions is
  zero.  As such, both $m^2$ and $\tilde{A}$ would be diagonal in the SCKM
  basis in the absence of the RH neutrino field, but in the presence
  of the see-saw mechanism off-diagonal elements are present in the
SCKM basis leading to LFV. Since the four relevent
  soft masses are degenerate at this point, 
  $m^2_{C^{5_1 5_2}} = m^2_{C^{5_1}_1} = m^2_{C^{5_1}_2} = m^2_{C^{5_1}_3}$, both
  D-term contributions are zero; $D^2_\theta = 0 = D^2_H$.
  
\item
  Point B is referred to as ``SUGRA''. With $X_S\neq X_T$ it
  represents typical flavour violation from the moduli fields,
  manifested as diagonal soft mass matrices not proportional to the 
unit matrix in the theory basis; this is
  the amount of flavour violation that would traditionally have been
  expected with no contribution from the $F_H$ or $F_\theta$ fields.
  This and point E are the only benchmark points investigated where
  $D_\theta \neq 0$.
  
\item
  Point C is referred to as ``FN flavour violation''.  It
  represents flavour violation from the Froggatt-Nielsen sector by
  itself, arising as a non-alignment of the trilinear soft terms
via the FN fields in the Yukawa operators,
without any contribution to flavour violation from traditional
  SUGRA effects, since $X_S=X_T$ as in point A.
  
\item
  Point D is referred to as ``Heavy Higgs flavour violation''.
  It represents flavour violation from the heavy Higgs sector, 
arising as a non-alignment of the trilinear soft terms, 
via the Heavy Higgs fields in the Yukawa operators,
without
  any contribution from either traditional SUGRA effects since
  $X_S=X_T$, or from FN fields since $F_\theta=0$.

\item
  Point E combines features of points B and C, resulting in
  Froggatt-Nielsen flavour violation from $X_\theta$, with SUGRA flavour
  violation from $X_S\neq X_T$. This is the only point where we see the
  flavour violation from the Froggatt-Nielsen fields and the $U(1)$ D-terms
  appearing at the same time. The numerical values for this point were
  obtained by taking the ratio from $X_S$ and $X_T$ for benchmark point B
  and applying it to benchmark point C.

\end{itemize}


\subsection{Varying $Y^e_{12}$ and $Y^e_{13}$}
\label{sec:varying-ye_12}


Normally, the chargino contribution to LFV dominates. Since the Feynman
diagram for this includes the left-handed sfermions, we would expect
the D-term corrections to the left-handed slepton mass matrix to
dominate the flavour violation. However, Model 2, as defined in
Section~\ref{sec:model-2}, is set up to have universal left-handed
charges, so the D-term correction from the breaking of $U(1)_F$ will
not contribute to flavour violation (except that it will either add or
remove some mass suppression). The D-term is limited in magnitude by
the difference of $m^2_\theta$ and $m^2_{\overline{\theta}}$, and although this
is not a strong correction to the soft masses, it can contribute
significantly to the lepton flavour violating branching ratios.

The difference in $\mu\rightarrow e\gamma$ between Model 1 and Model
2 is negligible for $Y^e_{12} = Y^e_{13}=0$. 
This should not be surprising, since
the texture zero coming from $Y^e_{12} = Y^e_{13}=0$ 
will yield small mixing angles,
resulting in small lepton flavour violation.

In order to get a picture of how great an effect the D-term
contributions could have on the soft masses, it is
necessary to examine a range of different values of
$Y^e_{12}$ and $Y^e_{13}$. An extra operator contribution was added to the
textures in Model 1 and Model 2, when compared to the model previously
studied \cite{KP0307091}, to allow for variations of the order
$Y^e_{12} \approx 10^{-3}$, for example.
The gives $\mathcal{O}(1)$ parameters $a''_{12}$
and $a'''_{12}$ for Models 1 and 2 respectively. To be precise $Y^e_{12} =
1.5\times10^{-3}$ corresponds to $a''_{12}$ and $a'''_{12} = 3.2$, and 
$Y^e_{13} = 1.5\times10^{-2}$ corresponds to $a_{13}=0.320$.


\subsection{Varying brane assignments for $\overline\theta$}
\label{sec:thetabar_brane_assignments}


The $\theta$ field is fixed to reside on the $C^{5_1 5_2}$ brane, but we
allow the brane assignment of $\overline\theta$ to
vary over the possibilities $C^{5_1 5_2}$, $C^{5_1}_1$, $C^{5_1}_2$
and $C^{5_1}_3$.  
This gives us D-terms that are calculable in each case, rather than being
free parameters.
The assignment of $\overline{\theta}$ to $C^{5_1 5_2}$,
which is the same assigmment as the $\theta$ field, implies that
the $U(1)_F$ D-term calculated in this case is zero.
The other possibilities, given in Eqs.~(\ref{eq:C^5_1 _1 D^2_theta} -
\ref{eq:C^5_1 _3 D^2_theta}), will highlight the contribution of the D-terms
to lepton flavour violation.


\subsection{Numerical procedure}


The code used to generate all the data here was based on SOFTSUSY
\cite{Allanach:2001kg}, which is a program that accurately calculates the spectrum
of superparticles in the MSSM. It solves the renormalisation group equations
with theoretical constraints on soft supersymmetry breaking terms provided
by the user. Successful radiative electroweak symmetry breaking is used as
a boundary condition, as are weak-scale
gauge coupling and fermion mass data (including one-loop finite MSSM
corrections). The program can also calculate a measure of fine-tuning. The 
program structure has, in this case, been adapted to the extension of the
MSSM considered in this paper. It is modified to include right-handed
neutrino fields, and thus non-zero neutrino masses and mixing angles, 
generated via the SUSY see-saw mechanism. It is also set up to include the
new D-term contributions considered herein, and to run
over a series of string assignments for the $\overline{\theta}$ field. We use
 $\tan \beta = 50$.

Electroweak symmetry breaking provides a significant constraint on the 
results. The breakdown of electroweak symmetry breaking was
responsible for the `spike' feature that was shown in the plots for 
benchmark points A and B in \cite{KP0307091}. 
For the data above the spike,
radiative electroweak symmetry breaking does not work properly, as
the Z-boson mass becomes tachyonic. In the present paper
such `bad' regions where electroweak symmetry breaking fails are
cut-off, however there is still a remnant of
the spike left, which is why one can see a slight rise at the ends of the plots
for our benchmark points A and B, as can be seen in
Section~\ref{sec:numerical results}.


\subsection{Numerical results}
\label{sec:numerical results}


We have now defined our two models, Model 1 and Model 2, and a set of
five benchmark points in Table~\ref{tab:benchmarks_defined} to examine
within them.  We have also set up what we will be varying apart from the
gravitino mass in these models -- the values of $Y^e_{12}$, 
$Y^e_{13}$ and the string 
assignment of $\overline\theta$ which gives different D-terms. We are now
in a position to present our results.
We shall focus on the branching ratios for $\mu \to e \gamma$ and
$\tau \to \mu \gamma$.
The branching ratio for $\tau\to e\gamma$ is not shown here as it does
not constrain us beyond those limits placed by $\mu \to e \gamma$ and
$\tau \to \mu \gamma$. The experimental limit for $\tau\to e\gamma$,
at $2.7\times10^{-6}$, is in fact far above the predicted rate for
this process at all examined parts of the parameter space.
In the following plots, we do not consider the $\overline{\theta}$
assignment to $C^{5_1}_3$ as this is exactly the same as $C^{5_1}_2$,
due to the degeneracy of the $X_{T_i}$. Were we to allow the $X_{T_i}$
to be non-degenerate, the phenomenological results of assigning
$\overline{\theta}$ to $C^{5_1}_2$ and $C^{5_1}_3$ would not be the
same. The detailed spectrum
will look different at each parameter point,
but the general trend is for the
physical masses to increase in
magnitude as the gravitino mass increases. 
Thus too-high gravitino masses
will start to reintroduce the fine-tuning problem resulting from the gluino
mass being too high \cite{Kane:1998im}, although we shall not
discuss the detailed spectrum here.

\begin{figure}[htbp]
\input{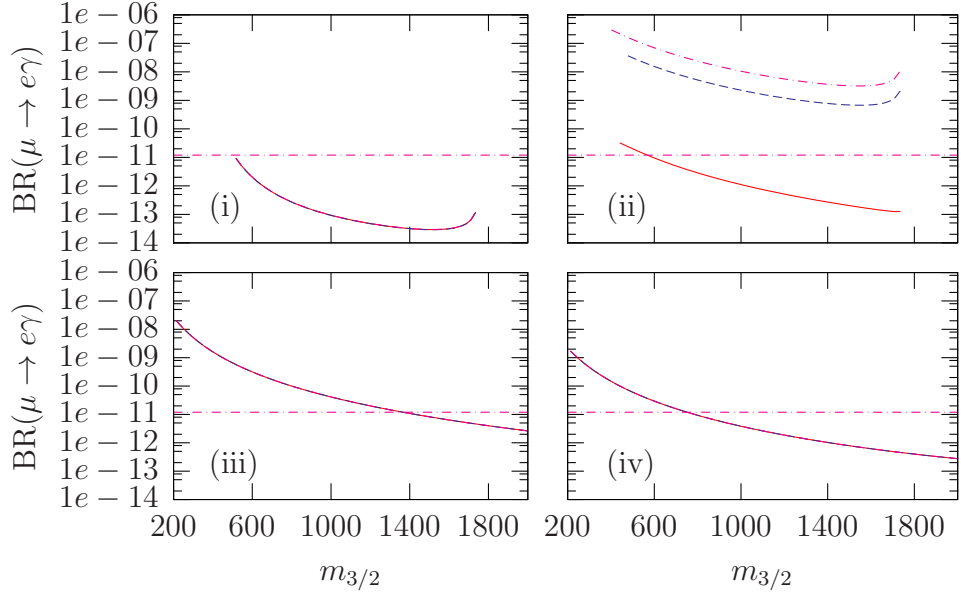} \centering
\caption{\small{Plots showing the Branching Ratio for $\mu \to e
\gamma$ for Model 1 with $Y^e_{12} = Y^e_{13} = 0$. Panel (i)
corresponds to benchmark point A, panel (ii) is for B, panel (iii) is
for C and panel (iv) is for D.  The $\overline{\theta}$ assignments
are shown with the separate lines: $C^{5_1 5_2}$ ({\em solid}),
$C^{5_1}_1$ ({\em dashed}), and $C^{5_1}_2$ ({\em dot-dash}). 
The solid curve corresponds to zero D-terms, and the other 
curves correspond to different models for the D-terms. 
The 2002 experimental limit \cite{Hagiwara:fs} is also given by the
horizontal line.}}
\label{fig:m32_meg_ABCD_model1}
\end{figure}

Figure~\ref{fig:m32_meg_ABCD_model1} shows numerical results for
$\mathrm{BR}(\mu\to e\gamma)$ for Model 1, plotted against the
gravitino mass $m_{3/2}$, where each of the four panels (i) --
(iv) correspond to each of the four benchmark points A -- D. 
\footnote{The solid line on each plot in
Figure~\ref{fig:m32_meg_ABCD_model1} corresponds to the solid
lines in Figure~1 of \cite{KP0307091}. However there were errors
in the code used to generate the previous data, and the corrected
rates shown here differ to the previous results by up to two
orders of magnitude. Furthermore, unlike the results in
\cite{KP0307091}, the results here do not exhibit a sharp spike
for benchmark points A and B. For the data above the spike,
radiative electroweak symmetry breaking does not work properly, as
the Z-boson mass becomes tachyonic. This was not realized in the
previous analysis.}
Panel (i) of Figure~\ref{fig:m32_meg_ABCD_model1} refers to
benchmark point A,
corresponding to minimum flavour violation, where the only source of
LFV is from the see-saw mechanism, which for Model 1 is well below
the experimental limit, shown as the horizontal dot-dash line.
Panel (ii) of
Figure~\ref{fig:m32_meg_ABCD_model1} refers to benchmark point B, 
with LFV arising from SUGRA, with the FN and heavy Higgs sources of LFV
switched off. In this case one can clearly distinguish the
additional contributions to LFV arising from the D-terms. This
makes benchmark point B the most phenomenologically interesting
for the purposes of this study. The differing contributions stem
from the $\overline{\theta}$ string assignments, which are shown
by the separate lines: $C^{5_1 5_2}$ ({\em solid}), $C^{5_1}_1$
({\em dashed}), and $C^{5_1}_2$ ({\em dot-dash}). The $C^{5_1
5_2}$ case shows the zero-D-term limit, where $\theta$ and
$\overline{\theta}$ are both intersection states, and hence
conspire to cancel out $D_\theta$ via their soft masses being
degenerate, $m^2_\theta - m^2_{\overline{\theta}} = 0$. The other
two locations for $\overline{\theta}$ then turn on the
$D_\theta$-term contributions. With the D-terms switched on, Model
1 is experimentally ruled out over all parameter space shown here.
Panel (iii) of
Figure~\ref{fig:m32_meg_ABCD_model1}, refering to benchmark point C, is
the Froggatt-Nielsen benchmark point, and for this case we see
that the experimental limit is satisfied for $m_{3/2}$ over
1400~GeV. Panel (iv) of
Figure~\ref{fig:m32_meg_ABCD_model1}, benchmark point D, shows
the heavy Higgs point, for which the experimental limit is
satisfied everywhere above 800~GeV. 
\footnote{Note that the predictions in this figure are
lower than the corresponding figure in \cite{KP0307091} due to the corrected 
values of $S,T_3$, as discussed.}

\begin{figure}[htbp]
\input{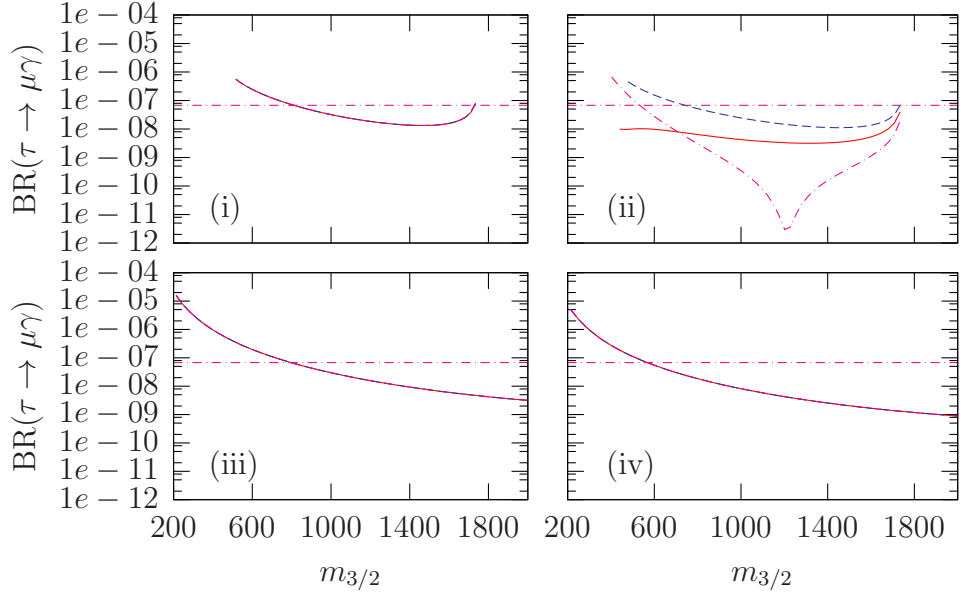} \centering
\caption{\small{Plots showing the Branching Ratio for $\tau
\to\mu\gamma$ for Model 1 with $Y^e_{12} = Y^e_{13} = 0$. Panel (i)
corresponds to benchmark point A, panel (ii) is for B, panel (iii) is
for C and panel (iv) is for D.  The $\overline{\theta}$ assignments
are shown with the separate lines: $C^{5_1 5_2}$ ({\em solid}),
$C^{5_1}_1$ ({\em dashed}), and $C^{5_1}_2$ ({\em dot-dash}).  
The solid curve corresponds to zero D-terms, and the other 
curves correspond to different models for the D-terms. 
The 2005 experimental limit \cite{Aubert:2005ye} is also given by the
horizontal line.}}
\label{fig:m32_tmg_ABCD_model1}
\end{figure}

Figure~\ref{fig:m32_tmg_ABCD_model1} shows analogous results for
$\mathrm{BR}(\tau\rightarrow\mu\gamma)$ for Model 1, plotted against
the gravitino mass $m_{3/2}$. All benchmark points come below the
experimental limit for a substantial amount of the parameter
space. The experimental limit here is more recent, and subsequently
much more stringent than the previous limit.  For these models in
which there is a large (2,3) element in the neutrino Yukawa matrix the
branching ratio for $\tau\to\mu\gamma$ is essentially as constraining
as that for $\mu\to e\gamma$, as first pointed out in
\cite{Blazek:2002wq}.  The D-term coupling to right-handed scalars has
a Yukawa mixing angle of order $\lambda^3$, compared to $\lambda^2$
for $\mu\rightarrow e\gamma$.  $\lambda \approx 0.22$ is the
Wolfenstein parameter, which contributes on an equal footing to
$\epsilon$ and $\delta$. So the right-handed sector is of equal
importance to the left-handed sector. We note that the see-saw effect
enters prominently in the left-handed sector, and by considering
Eqs.~(\ref{eq:m^2_QL}) and (\ref{eq:m^2_LL}) for the soft scalar
masses in the $(2-3)$ sector for $\tau \to \mu \gamma$, one can show
that there is little effect coming from the D-terms coupling to
left-handed scalars, since we have universal family charges for the
left-handed $(2-3)$ sector, $q_{L_2} = q_{L_3}$. For the right-handed
scalars, however, the D-terms do play an important part and can have
rather interesting and surprising effects, as we now
discuss in some detail.

The solid line in panel (ii) of Figure~\ref{fig:m32_tmg_ABCD_model1}
for the $C^{5_1 5_2}$ string assignment of $\overline{\theta}$
has zero contribution from the $U(1)_F$ D-terms, and shows just the
effect of non-minimal SUGRA. This actually suppresses the flavour
violation arising from the see-saw effect alone, showing an
interesting cancellation between the LFV from the see-saw
mechanism and the LFV from the non-universal D-terms.
On the other hand the dashed line in
panel (ii) for the $C^{5_1}_1$ case is very similar to the see-saw
scenario of benchmark point A shown in panel (i). This is due to the
D-terms actually conspiring to restore universality in the scalar
masses, turning non-minimal SUGRA back into the minimal form. 
One can easily see this by applying Eq.~(\ref{eq:C^5_1 _1 D^2_theta}) to
Eqs.~(\ref{eq:m^2_UR}) - (\ref{eq:m^2_NR}) for the right-handed scalar
mass matrices, as the D-terms bring in an equal but opposite effect to
the non-universal effects from SUGRA, and subsequently force the mass
matrices to become universal. It is an amazing consequence of this
string assignment for $\overline{\theta}$ that in this model the
effects of the non-universal $U(1)_F$ D-terms can exactly cancel the
effects of the non-universal SUGRA for the branching ratio of $\tau
\to \mu \gamma$, leading to universal scalar mass matrices, even with
SUGRA turned on.  This is a string effect that directly affects the
amount of flavour violation predicted in this scenario. For the
$C^{5_1}_2$ case shown by the dot-dash line, applying
Eq.~(\ref{eq:C^5_1 _2 D^2_theta}) to the right-handed scalar mass
matrices Eqs.~(\ref{eq:m^2_UR}) - (\ref{eq:m^2_NR}) shows that the
D-terms in this case actually enhance the effect of non-minimal SUGRA,
causing the scalar mass matrices to become even more non-universal.
One can understand this purely right-handed effect by considering the
different mass insertion diagrams for the left- and right-handed
sectors. The left-handed sector involves charginos, whereas the
right-handed sector only involves neutralinos, so the right-handed
masses scale differently with the gravitino mass as compared to the
left-handed masses. Thus we do not have a universal mass scaling
between the left- and right-handed sectors. This leads to the observed
smooth cancellation of flavour violation between the two competing
contributions from the see-saw mechanism and from SUGRA with
additional $U(1)_F$ D-terms.




\begin{figure}[htbp]
\input{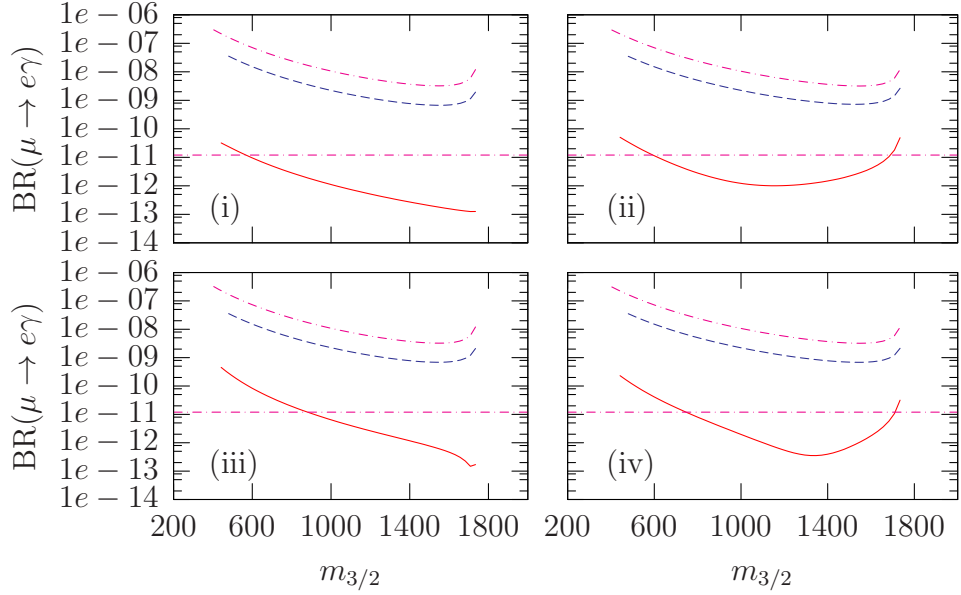}
\centering
\caption{\small{Plots showing the Branching Ratio for $\mu \to e \gamma$ for
benchmark point B for Model 1 only.
Panel (i) has $Y^e_{12} = 0$ and $Y^e_{13} = 0$.
Panel (ii) has $Y^e_{12} = 1.5\times10^{-3}$ and $Y^e_{13} = 0$.
Panel (iii) has $Y^e_{12} = 0$ and $Y^e_{13} = 1.5\times10^{-2}$.
Panel (iv) has $Y^e_{12} = 1.5\times10^{-3}$ and $Y^e_{13} = 1.5\times10^{-2}$.
The $\overline{\theta}$ assignments are shown with the separate lines:
$C^{5_1 5_2}$ ({\em solid}), $C^{5_1}_1$ ({\em dashed}), and $C^{5_1}_2$
({\em dot-dash}).
The solid curve corresponds to zero D-terms, and the other 
curves correspond to different models for the D-terms. 
The 2002 experimental limit \cite{Hagiwara:fs} is also given by the horizontal line.}}
\label{fig:m32_meg_B_model1_Y}
\end{figure}

Figure~\ref{fig:m32_meg_B_model1_Y} shows benchmark point B for Model
1.  The four panels show the $Y^e_{12}$ and $Y^e_{13}$ electron Yukawa
elements being turned on and
off. The results for 
Figure~\ref{fig:m32_meg_B_model1_Y}(i) are for $Y^e_{12} = 0$ and
$Y^e_{13} = 0$. This is the same as in panel (ii) of
Figure~\ref{fig:m32_meg_ABCD_model1}, and is the base from which we
start.  Panel (ii) of Figure~\ref{fig:m32_meg_B_model1_Y} has $Y^e_{12}
= 1.5\times10^{-3}$ and $Y^e_{13} = 0$, so we can clearly see the
effect of turning $Y^e_{12}$ on. It only affects the $C^{5_1 5_2}$
line, as the D-terms dominate over this effect for the other two
string assignments.  Panel (iii) of
Figure~\ref{fig:m32_meg_B_model1_Y} uses $Y^e_{12} = 0$ and $Y^e_{13}
= 1.5\times10^{-2}$, highlighting the effect of just $Y^e_{13}$ alone.
Again the zero D-term line of $C^{5_1 5_2}$ is the only one that is
sizably affected by this change in Yukawa texture.  Panel (iv) of
Figure~\ref{fig:m32_meg_B_model1_Y} shows the effect of turning on
both Yukawa elements: $Y^e_{12} = 1.5\times10^{-3}$ and $Y^e_{13} =
1.5\times10^{-2}$.  We see that the shape of the solid line is
determined by both Yukawa textures -- they seem to have an equal
impact on it.

\begin{figure}[htbp]
\input{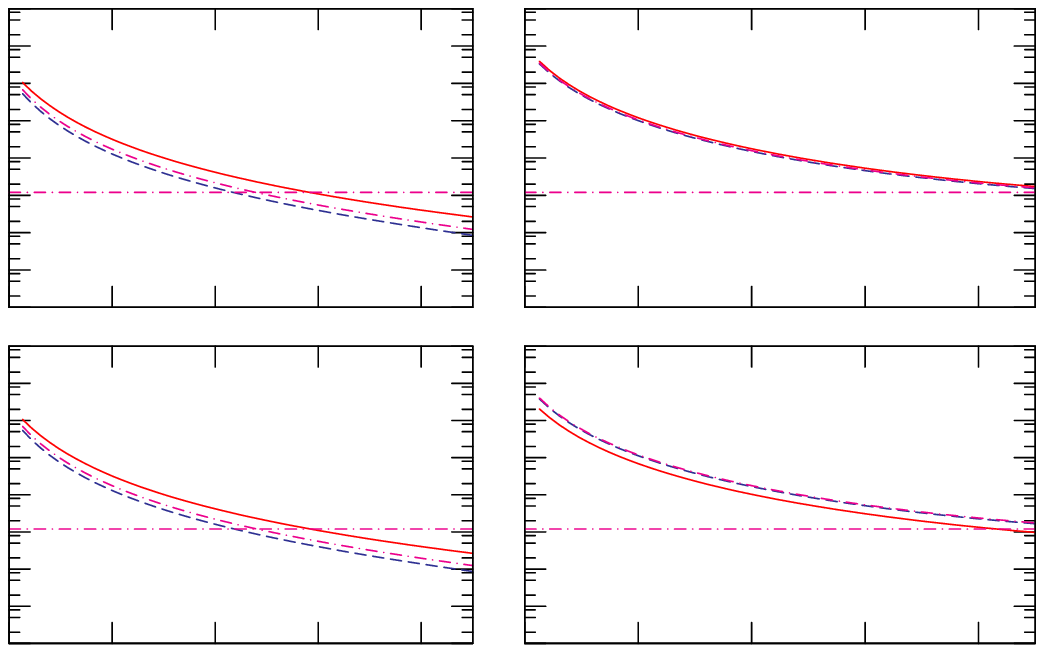}
\centering
\caption{\small{Plots showing the Branching Ratio for $\mu \to e \gamma$ for
benchmark point E for Model 1 only.
Panel (i) has $Y^e_{12} = 0$ and $Y^e_{13} = 0$.
Panel (ii) has $Y^e_{12} = 1.5\times10^{-3}$ and $Y^e_{13} = 0$.
Panel (iii) has $Y^e_{12} = 0$ and $Y^e_{13} = 1.5\times10^{-2}$.
Panel (iv) has $Y^e_{12} = 1.5\times10^{-3}$ and $Y^e_{13} = 1.5\times10^{-2}$.
The $\overline{\theta}$ assignments are shown with the separate lines:
$C^{5_1 5_2}$ ({\em solid}), $C^{5_1}_1$ ({\em dashed}), and $C^{5_1}_2$
({\em dot-dash}).
The solid curve corresponds to zero D-terms, and the other 
curves correspond to different models for the D-terms. 
The 2002 experimental limit \cite{Hagiwara:fs} is also given by the horizontal line.}}
\label{fig:m32_meg_E_model1_Y}
\end{figure}

\begin{figure}[htbp]
\input{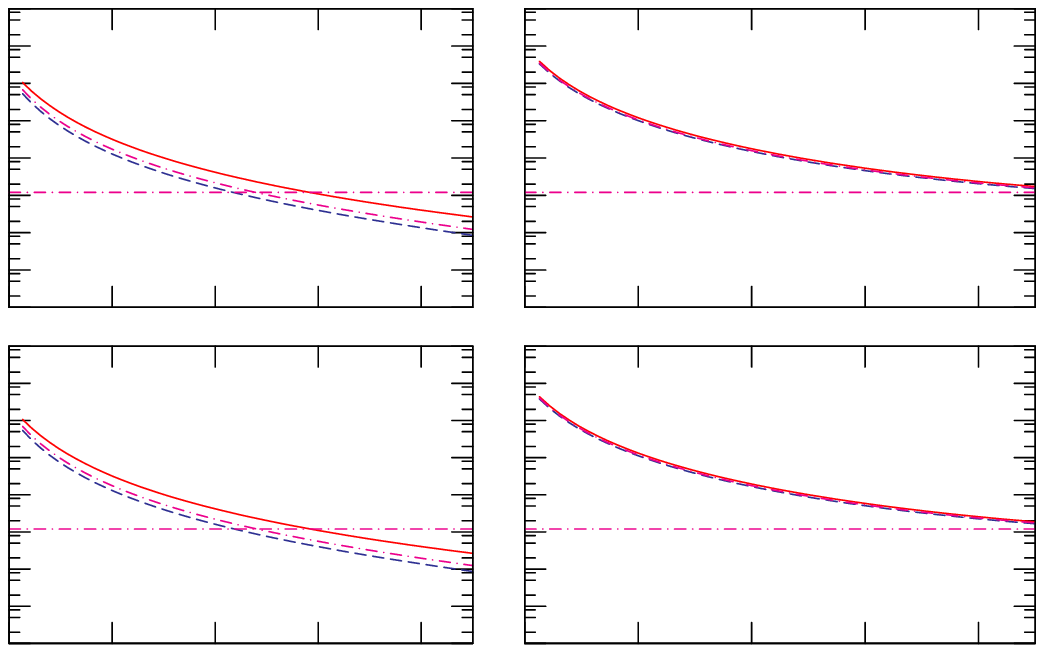}
\centering
\caption{\small{Plots showing the Branching Ratio for $\mu \to e \gamma$ for
benchmark point E for Model 2 only.
Panel (i) has $Y^e_{12} = 0$ and $Y^e_{13} = 0$.
Panel (ii) has $Y^e_{12} = 1.5\times10^{-3}$ and $Y^e_{13} = 0$.
Panel (iii) has $Y^e_{12} = 0$ and $Y^e_{13} = 1.5\times10^{-2}$.
Panel (iv) has $Y^e_{12} = 1.5\times10^{-3}$ and $Y^e_{13} = 1.5\times10^{-2}$.
The $\overline{\theta}$ assignments are shown with the separate lines:
$C^{5_1 5_2}$ ({\em solid}), $C^{5_1}_1$ ({\em dashed}), and $C^{5_1}_2$
({\em dot-dash}).
The solid curve corresponds to zero D-terms, and the other 
curves correspond to different models for the D-terms. 
The 2002 experimental limit \cite{Hagiwara:fs} is also given by the horizontal line.}}
\label{fig:m32_meg_E_model2_Y}
\end{figure}

Figure~\ref{fig:m32_meg_E_model1_Y} shows benchmark point E, which combines
the features of benchmark points B and C, thus both $U(1)$ D-term and
Froggatt-Nielsen (FN) flavour violation appear in the predicted branching
ratios shown in this figure. There is some interesting
interplay between the FN F-terms in the benchmark point C region of
parameter space, and the D-terms in benchmark point B, and benchmark point E
is designed to show this difference. In Figure~\ref{fig:m32_meg_E_model1_Y} 
the results are shown for 
Model 1 with both $Y^e_{12}$ and $Y^e_{13}$ Yukawa elements on
and off. The shape of the curves are as in benchmark point C due to $X_\theta$
being turned on, and the results are numerically similar to those of benchmark
point C, showing that the dominant contribution comes from Froggatt-Nielsen
flavour violation. However, as in benchmark point B, the different
D-terms corresponding to different $\overline{\theta}$
assignments leads to noticeable shifts in the results.
In panel (i) with $Y^e_{12} = 0$ and $Y^e_{13} = 0$
it is seen that the presence of non-zero D-terms actually
reduces the LFV rate somewhat compared to the solid curve
with zero D-terms, corresponding to a region of parameter space where there is
some cancellation between the flavour violation from the Froggat-Nielsen
fields and that caused by the $U(1)_F$ D-terms. 
The other panels show variation of the Yukawa elements as in
Figure~\ref{fig:m32_meg_B_model1_Y},
for example panel (iv) corresponds to both 
$Y^e_{12}$ and $Y^e_{13}$ Yukawa elements being non-zero.
Note that the solid curve in panel (iv) 
is slightly lower than the solid curve in panel (ii),
showing that sometimes a non-zero Yukawa coupling can reduce LFV.

Figure~\ref{fig:m32_meg_E_model2_Y} shows the effects of the Yukawa
elements for benchmark point E using Model 2, where Model 2
has the same Yukawa structure as Model 1, but has the feature that
the left-handed family charges are the same for all three families,
resulting in universal D-terms, at least in the left-handed sector.
Comparing Figure~\ref{fig:m32_meg_E_model2_Y} to 
Figure~\ref{fig:m32_meg_E_model1_Y}, we see that 
in panel (i) of both figures with $Y^e_{12} = 0$ and $Y^e_{13} = 0$
there is no obserable difference between the predictions 
of the two models. However comparing panels (iv) of
Figure~\ref{fig:m32_meg_E_model2_Y} and Figure~\ref{fig:m32_meg_E_model1_Y}
we see that with non-zero $Y^e_{12}$ and $Y^e_{13}$ Model 2 has
the effect of reducing the LFV resulting from the D-terms.

We could have presented similar plots for benchmark point D, with a new
point F,\footnote{The values for the $X$ parameters for benchmark
point F would be
\begin{equation} \nonumber
\begin{array}{cccccc}
X_S & X_T & X_H & X_{\overline{H}} & X_\theta & X_{\overline{\theta}} \\
0.290 & 0.264 & 0.595 & 0.595 & 0.000 & 0.000
\end{array} .
\end{equation}}
which is an amalgamation of points B and D, but since this
produces the same kind of results as Figure~\ref{fig:m32_meg_E_model1_Y},
we did not present the results.


\section{Conclusions} 
\label{sec:conclusions} 


We have catalogued and quantitatively studied the importance of
all the different sources of LFV present
in a general non-minimal SUGRA framework, including the 
effects of gauged family symmetry. 
We have discussed five different sources of LFV in such models:
see-saw induced LFV arising from the running effects of right-handed neutrino
fields; supergravity induced LFV
due to the non-universal structure of the supergravity
model; FN (Higgs) flavour violation, due to the F-terms associated with FN 
(Higgs) fields developing \vevs{}, and contributing in a non-universal way 
to the soft trilinear terms; D-term flavour violation, where the D-term mass
correction from the breaking of the Abelian family symmetry drives the
scalar mass matrices to be non-universal; and finally the
effects of different choices of Yukawa textures on LFV.

In order to quantify the importance of the different effects
we investigated these disparate souces of LFV numerically, 
both in isolation and
in association with one another, within a particular SUGRA 
model based on a type I string-inspired Pati-Salam model 
with an Abelian family symmetry, which has a sufficiently rich structure
to enable all of the effects to studied within a single framework.
Within this framework
we derived the soft supersymmetry breaking terms, including the effect
of the D-terms associated with breaking the family symmetry. 
For these models the D-terms are calculable, but are
model dependent, depending on a particular choice of string assignment
for the FN fields, and in particular the D-terms are only non-zero
for the non-universal SUGRA models. We have performed
a detailed numerical analysis of the five sources of LFV using five
benchmark points designed to highlight the particular effects,
and we have explored the effect of the variation of Yukawa texture elements 
on the results.

The most striking conclusion is how dangerously large the calculable D-term
contribution to flavour violation can be, at least for the class
of models studied. However it should be emphasised that
while the D-terms are calculable in these models they are
also model dependent, and it is always possible to 
simply switch off the D-terms by selecting
the $\overline{\theta}$ field to have the same string assignment
as the ${\theta}$ intersection state. However
other choices will lead to non-zero but calculable D-terms,
which can be dangerously large, or can massively suppress flavour violation.
For example the curves with non-zero D-terms 
in Figure~\ref{fig:m32_meg_B_model1_Y} all exceed the
experimental limit for the Branching Ratio for $\mu \to e \gamma$,
showing that D-term effects have the potential to greatly exceed the
other contributions from SUGRA, the see-saw mechanism, FN and Higgs, depending
on the choice of Yukawa textures.
However in some cases the D-terms generated by breaking the $U(1)_F$ family
symmetry can also suppress the Branching Ratio for $\mu \to e \gamma$,
as shown in panels (i) and (iii) of
Figures~\ref{fig:m32_meg_E_model1_Y} and \ref{fig:m32_meg_E_model2_Y}.

Another notable feature is the effect of Yukawa texture on the
results. 
The Yukawa texture has $Y^e_{12} = Y^e_{13} = 0$, and we have shown
that turning on 
non-zero values of these Yukawa couplings can greatly enhance the
branching ratio for $\mu\to e\gamma$ almost arbitarily.
The reason is that the rotations to the SCKM basis are 
controlled by these Yukawa elements and the larger these
rotations the larger will be the off-diagonal soft masses in the SCKM basis.
The non-zero magnitudes of $Y^e_{12}$ and $Y^e_{13}$ 
were therefore chosen to be large enough
to show the variations in the branching ratios of the different
models, but small enough to keep within the currently
experimentally allowed range. 

In this paper we have worked with
a particular Yukawa texture in which there is a large (2,3) element in the
neutrino Yukawa matrix leading to large see-saw induced LFV and a 
branching ratio for $\tau\to\mu\gamma$ which is 
as constraining as that for $\mu\to e\gamma$ \cite{Blazek:2002wq}.
However we have seen that in some cases the D-terms can 
lead to a large suppression of the rate for particular 
values of $m_{3/2}$, as seen 
in panel (ii) of Figure~\ref{fig:m32_tmg_ABCD_model1}.
For other cases the effects of the non-universal $U(1)_F$ D-terms 
can exactly cancel the effects of the non-universal SUGRA model 
leading to universal scalar mass matrices, thereby restoring 
universality even for a non-minimal SUGRA model.
Such effects are only possible 
in certain string set-ups and thus LFV provides 
an observable signal which may discriminate between different
underlying string models.

In conclusion, we have seen that within realistic non-minimal
supergravity models there can be several important
effects leading to much larger LFV than in the case usually
considered in the literature of minimum flavour violation
corresponding to just mSUGRA and the see-saw mechanism, 
and considered here as benchmark point A. We find that the D-term
contributions are generally dangerously large, but in certain cases
such contributions can lead to a dramatic suppression of LFV rates,
for example by cancelling the effect of the see-saw induced
LFV in $\tau \to \mu \gamma$ models with lop-sided textures.
In the class of string models considered here we find the surprising result
that the D-terms can sometimes serve to restore 
universality in the effective non-minimal supergravity theory.
Thus D-terms can give very large and very
surprising effects in LFV processes.
In general there
will be a panoply of different sources of LFV in realistic
non-minimal SUGRA models, and we have explored the relative
importance of some of them within a particular framework.
The results here only serve to heighten the expectation
that LFV processes such as $\mu \to e \gamma$ and
$\tau\to\mu\gamma$ may be observed soon, although it is
clear from our results that the precise theoretical
interpretation of such signals will be more non-trivial
than is apparent from many previous studies in the literature.


\vskip 0.1in
\noindent
{\large {\bf Acknowledgements}}\\
J.H. thanks PPARC for a studentship.


\newpage
\appendix

\numberwithin{equation}{section} 



\section{Parameterised trilinears for the 42241 model}
\label{sec:param-tril-42241}


We here write the general form of the trilinear parameters $A_{ijk}$ assuming
nothing about the form of the Yukawa matricies.
\begin{eqnarray}
  \nonumber
  A_{C_1^{5_1} C^{5_1 5_2} C^{5_1 5_2}}
  &=&
  \sqrt{3} m_{3/2}
  \left\{
    X_S
      \left[
        1 +
        \left(S + \overline{S}\right)\partial_S \ln Y_{abc} 
      \right]
    \right. 
  \\
  &
  \nonumber
  & 
  {} + X_{T_1}
    \left[
      -1 + \left(T_1 + \overline{T}_1 \right)\partial_{T_1} \ln Y_{abc}
    \right]
  \\
  &&
  \nonumber
  {} + X_{T_2}
  \left[
    -1 + \left(T_2 + \overline{T}_2 \right) \partial_{T_2} \ln Y_{abc}
  \right]
  \\
  \nonumber
  &&
  {} + X_{T_3}
    \left(T_3 + \overline{T}_3 \right)\partial_{T_3} \ln Y_{abc}
  \\
  \nonumber
  &&
  {} + X_H
    \left(S + \overline{S}\right)^{\frac{1}{2}} H \partial_H \ln Y_{abc}
    \\
  &&
  \nonumber
  {} + X_{\overline{H}}
      \left(T_3 + \overline{T}_3\right)^\frac{1}{2}\overline{H} \partial_{\overline{H}}
      \ln Y_{abc}
      \\
  \label{eq:trilinear_one_42241}
  &&
  \left.
    {} + X_\theta
    \left(S+\overline{S}\right)^\frac{1}{4}
    \left(T_3+\overline{T}_3\right)^\frac{1}{4}
    \theta \partial_\theta \ln Y_{abc}
  \right\}
\end{eqnarray}

\begin{eqnarray}
  \nonumber
  A_{C_1^{5_1} C_3^{5_1} C^{5_1 5_2}}
  &=&
  \sqrt{3} m_{3/2}
  \left\{
    X_S
      \left[
        \frac{1}{2} +
        \left(S + \overline{S}\right)\partial_S \ln Y_{abc} 
      \right]
    \right. 
  \\
  &&
  \nonumber
  {} + X_{T_1} 
    \left[
      -1 + \left(T_1 + \overline{T}_1 \right)\partial_{T_1} \ln Y_{abc}
    \right]
  \\
  &&
  \nonumber
  {} + X_{T_2}
    \left(T_2 + \overline{T}_2 \right) \partial_{T_2} \ln Y_{abc}
  \\
  &&
  \nonumber
  {} + X_{T_3}
    \left[
      - \frac{1}{2}
      \left(T_3 + \overline{T}_3 \right)\partial_{T_3} \ln Y_{abc}
    \right]
  \\
  \nonumber
  &&
  {} + X_H
    \left(S + \overline{S}\right)^{\frac{1}{2}}H \partial_H \ln Y_{abc}
  \\
  \nonumber
  &&
  {} +
    X_{\overline{H}}
    \left(T_3 + \overline{T}_3\right)^\frac{1}{2}\overline{H} \partial_{\overline{H}}
    \ln Y_{abc}
    \\
  \label{eq:trilinear_two_42241}
    &&
    \left.
    {} + X_\theta
    \left(S+\overline{S}\right)^\frac{1}{4}
    \left(T_3+\overline{T}_3\right)^\frac{1}{4}
    \theta \partial_\theta \ln Y_{abc}
  \right\}
\end{eqnarray}

\begin{eqnarray}
  \nonumber
  A_{C_1^{5_1} C_2^{5_1} C^{5_1 5_2}}
  &=&
  \sqrt{3} m_{3/2}
  \left\{
    X_S
      \left[
        \frac{1}{2} +
        \left(S + \overline{S}\right)\partial_S \ln Y_{abc} 
      \right]
    \right. 
  \\
  &&
  \nonumber
  {} + 
  X_{T_1}
    \left[
      -1 + \left(T_1 + \overline{T}_1 \right)\partial_{T_1} \ln Y_{abc}
    \right]
  \\
  &&
  \nonumber
  {} + X_{T_2} 
  \left[
    -1 + \left(T_2 + \overline{T}_2 \right) \partial_{T_2} \ln Y_{abc}
  \right]
  \\
  \nonumber
  &&
    {} +
    X_{T_3}
    \left[
      \frac{1}{2}  
      \left(T_3 + \overline{T}_3 \right)\partial_{T_3} \ln Y_{abc}
    \right]
  \\
  \nonumber
  &&
  {} + X_H
    \left(S + \overline{S}\right)^{\frac{1}{2}}H \partial_H \ln Y_{abc}
  \\
  &&
  \nonumber
    {} + X_{\overline{H}}
    \left(T_3 + \overline{T}_3\right)^\frac{1}{2}\overline{H} \partial_{\overline{H}}
    \ln Y_{abc}
    \\
  &&
  \label{eq:trilinear_three_42241}
  \left.
    {} + X_\theta
    \left(S+\overline{S}\right)^\frac{1}{4}
    \left(T_3+\overline{T}_3\right)^\frac{1}{4}
    \theta \partial_\theta \ln Y_{abc}
  \right\}
\end{eqnarray}

\begin{eqnarray}
  \nonumber
  A_{C_1^{5_1} C_2^{5_1} C_3^{5_1}}
  &=&
  \sqrt{3} m_{3/2}
  \left\{
    X_S
        \left(S + \overline{S}\right)\partial_S \ln Y_{abc} 
    \right. 
  \\
  &&
  \nonumber
  {} + X_{T_1}
    \left[
      -1 + \left(T_1 + \overline{T}_1 \right)\partial_{T_1} \ln Y_{abc}
    \right]
  \\
  &&
  \nonumber
  {} + X_{T_2}
    \left(T_2 + \overline{T}_2 \right) \partial_{T_2} \ln Y_{abc}
  \\
  \nonumber
  &&
  {} + X_{T_3}
    \left(T_3 + \overline{T}_3 \right)\partial_{T_3} \ln Y_{abc}
  \\
  \nonumber
  &&
  {} + X_H
    \left(S + \overline{S}\right)^{\frac{1}{2}}H \partial_H \ln Y_{abc}
  \\
  \nonumber
  &&
    {} +
    X_{\overline{H}}
    \left(T_3 + \overline{T}_3\right)^\frac{1}{2}\overline{H} \partial_{\overline{H}}
    \ln Y_{abc}
  \\
  \label{eq:trilinear_four_42241}
    &&
    \left.
    {} + X_\theta
    \left(S+\overline{S}\right)^\frac{1}{4}
    \left(T_3+\overline{T}_3\right)^\frac{1}{4}
    \theta \partial_\theta \ln Y_{abc}
  \right\}
\end{eqnarray}



\section{Derivation of the D-terms}
\label{sec:deriv D-terms}


What follows is a full derivation of $D_H$ and $D_{\theta}$ from the
superpotential of the $U(1)_F$ extended supersymmetric Pati-Salam 42241
model, as detailed in Section~\ref{sec:442-pati-salam}.

The relevent parts of the superpotential for the 42241 model are those
concerning the Higgs and Froggatt-Nielsen fields which have different gauge
singlets\footnote{Note that these must still have the same quantum numbers
as they are both singlets, and therefore in the same representation of the
gauge group.}.
\begin{equation}
W = S\lambda_S (\overline{H}H-M^2_H) + S'\lambda_{S'} (\overline{\theta}\theta
-M^2_\theta) ,
\end{equation}
where the $M_H$ and $M_\theta$ are GUT scale masses associated with the
Higgs and Froggatt-Nielsen vevs respectively. We have assumed that
the heavy Higgs  develop vevs along the neutrino directions only, such that
\begin{equation}
\langle\overline{H}\rangle = \overline{H}_\nu = \overline{H} \quad ; \quad
\langle H\rangle = H_\nu = H .
\end{equation}

This is because charged objects gaining vevs would break their charge group
at the GUT scale, causing problems\footnote{A colour charged object, for example,
would imply that QCD is broken at the GUT scale, which would lead to a very
massive gluon.} which the neutral components avoid. Similarly, the
Froggatt-Nielsen vevs are concisely written as
\begin{equation}
\langle\overline{\theta}\rangle = \overline{\theta} \quad ; \quad
\langle\theta\rangle = \theta .
\end{equation}

The F-terms associated with the singlet fields are
\begin{equation}
\label{eq:F_S-term}
|F_S|^2 = |\lambda_S(\overline{H}H -M^2_H)|^2
\end{equation}
and
\begin{equation}
\label{eq:F_S'-term}
|F_{S'}|^2 = |\lambda_{S'}(\overline{\theta}\theta -M^2_\theta)|^2 .
\end{equation}

We use these to form our Higgs potential,
\begin{equation}
V_H = V_D + V_F + V_{\mathrm{soft}} .
\end{equation}

The F-term potential is trivially obtained from Eqs.~(\ref{eq:F_S-term}) and
(\ref{eq:F_S'-term}), and the soft terms are simply written down with
mass-squared terms for each of the soft SUSY-breaking scalar masses associated
with the Higgs and FN vevs, as we will later see. The D-term potential
takes a little more work, so we shall cover that here. The general form for
the D-term potential is
\begin{equation}
\label{eq:general-V_D}
V_D = \frac{1}{2} g^2_F D^1_F D^1_F +\frac{1}{2} g^2_{2R} \sum\limits^3_{a=1}
D^a_{2R} D^a_{2R} + \frac{1}{2}g^2_4 \sum\limits^{15}_{m=1} D^m_4 D^m_4 ,
\end{equation} 
where $g_F$ is the gauge coupling for $U(1)_F$, $g_{2R}$ is for $SU(2)_R$,
and $g_4$ is for $SU(4)_c$.

We focus on the $a=3$ and $m=15$ contributions to Eq.~(\ref{eq:general-V_D})
which involve the
\begin{equation}
\label{eq:SU(2)_R-generator}
\tau^3_R = \mathrm{diag}\left( \frac{1}{2} , -\frac{1}{2} \right)
\end{equation}
and
\begin{equation}
\label{eq:SU(4)-generator}
T^{15}_4 = \sqrt{\frac{3}{2}}\mathrm{diag} \left( \frac{1}{6} , \frac{1}{6} , 
\frac{1}{6} , -\frac{1}{2} \right) ,
\end{equation}
the diagonal generators of the $SU(2)_R$ and $SU(4)$ groups. There is no sum
in the $U(1)$ part of Eq.~(\ref{eq:general-V_D}) because this group is
the unit matrix {\bf 1}, so the only generator is
\begin{equation}
\label{eq:U(1)_F-generator}
T^1_F = {\bf 1}q_F .
\end{equation}
where $q_F$ is the charge of the $U(1)_F$ group for each field. When
applying Eqs.~(\ref{eq:SU(2)_R-generator}) and (\ref{eq:SU(4)-generator})
to conjugate fields we have to complex conjugate the generator and
multiply it by $-1$, but for the $U(1)_F$ group we just use
Eq.~(\ref{eq:U(1)_F-generator}) where it is known that the $q_F$
charges are different for right-handed fields.

Using Table~\ref{tab:particle_content_42241} we find that $D^1_F$, $D^3_{2R}$,
and $D^{15}_4$ are given by
\begin{eqnarray}
\label{eq:D^1_F in terms of fields}
D^1_F &=& \overline{H}^\dagger (q_F)_{\overline{H}} \overline{H}
+ H^\dagger (q_F)_H H + \overline{\theta}^\dagger (q_F)_{\overline{\theta}}
\overline{\theta} + \theta^\dagger (q_F)_{\theta} \theta \nonumber \\ & &
+ \overline{F}^\dagger_i (q_F)_{\overline{F}_i} \overline{F}^j \delta^j_i
+ F^\dagger_i (q_F)_{F_i}F^j \delta^j_i + h^\dagger (q_F)_h h
\\
\label{eq:D^3_2R in terms of fields}
D^3_{2R} &=& \overline{H}^\dagger (-\tau^{3*}_R) \overline{H}
+ H^\dagger (\tau^3_R) H + \overline{F}^\dagger (-\tau^{3*}_R) \overline{F}
+ h^\dagger (\tau^3_R) h
\\
\label{eq:D^15_4 in terms of fields}
D^{15}_4 &=& \overline{H}^\dagger (-T^{15*}_4) \overline{H}
+ H^\dagger (T^{15}_4) H  + \overline{F}^\dagger (-T^{15*}_4) \overline{F}
+ F^\dagger (T^{15}_4) F.
\end{eqnarray}

In Eq.~(\ref{eq:D^1_F in terms of fields}), $i,j \in \{1,2,3\}$ are family
indices. We used $\delta^j_i$ to pick out the trace of the outer products
$F^\dagger_i F^j$ and $\overline{F}^\dagger_i \overline{F}^j$, thereby giving us
the dot product.

The scalar components of the left-handed matter superfield $F$ are $q$ and
$l$. The scalar components of the right-handed matter superfield $\overline{F}$
are $u^c$, $d^c$, $\nu^c$, and $e^c$. The tensorial conventions are shown in
\cite{King:2000vp} for $D^3_{2R}$ and $D^{15}_4$. For $D^1_F$, all diagonal
elements are equal to unity for the generator, so the tensor notation is trivial.

Now we square $D^1_F$, $D^3_{2R}$, and $D^{15}_4$ (as they are squared
in the D-term potential), then consider the cross terms, as we are
only interested in those terms that look like
$(mass)^2\cdot(field)^2$. The $(mass)^2$ terms come from the vevs of
the heavy Higgs and Froggatt-Nielsen fields $\overline{H}^2, H^2,
\overline{\theta}^2, \theta^2$. The $(field)^2$ terms come from
$|F_i|^2, |\overline{F}_i|^2$, where the $i$ is a family index running
from 1 to 3.

In Eqs.~(\ref{eq:m^2_QL}) to (\ref{eq:m^2_hd}) we have defined $D^2_H$ and
$D^2_\theta$ to be

\begin{eqnarray}
\label{eq:D^2_H initial}
D^2_H &=& \frac{1}{8} (\overline{H}^2 -H^2)\\ \label{eq:D^2_theta initial}
D^2_\theta &=& - q_\theta (\overline{\theta}^2 -\theta^2)
-q_H(\overline{H}^2 - H^2)  ,
\end{eqnarray}

where $q_\theta$ is defined to be $-1$ and $q_H$ is defined to be $-q_{R3}$, thus
$q_H = \frac{5}{6}$ for Model 1 and $q_H = 1$ for Model 2.

By considering the charge structure in Table~\ref{tab:charges_both_models}
we can clearly see how the charge structure effects the additional $U(1)_F$
D-term contributions to the sparticle mass matrices in 
Eqs.~(\ref{eq:m^2_QL})-(\ref{eq:m^2_hd}).

We have followed \cite{King:2000vp} in choosing our designation of $D^2_H$,
but $D^2_\theta$ is new. Working explicitly within Model 1, where
$q_H = \frac{5}{6}$, we shall proceed to rewrite these D-terms as functions
of the soft SUSY breaking masses and gauge couplings.
So we arrive at our D-term potential, which, when put together with $V_F$ and
$V_{\mathrm{soft}}$, forms our Higgs potential:
\begin{eqnarray}
V_{\mathrm{Higgs}} &=& \frac{1}{2} g^2_F \left[ \frac{25}{36} (\overline{H}^2
-H^2)^2 +(\overline{\theta}^2 -\theta^2)^2 -\frac{5}{3}(\overline{H}^2
-H^2)(\overline{\theta}^2 -\theta^2)\right] \nonumber \\ & &
+\frac{1}{8} (g^2_{2R} +\frac{3}{2} g^2_4)(\overline{H}^2 -H^2)^2
+\lambda^2_S(\overline{H}H -M^2_H)^2 +\lambda^2_{S'}
(\overline{\theta}\theta -M^2_\theta)^2 \nonumber \\ & &
+m^2_{\overline{H}} \overline{H}^2 +m^2_H H^2 +m^2_{\overline{\theta}}
\overline{\theta}^2 +m^2_\theta \theta^2 ,
\end{eqnarray}
where $V_D$ are the terms multiplied by gauge coupling factors, $V_F$ are
the terms multiplied by the dilaton lambdas, and $V_{\mathrm{soft}}$ are the
last four terms multiplying the TeV scale soft SUSY breaking scalar masses.
$M_H$ and $M_\theta$ are GUT scale masses.

To find the form of the D-terms, we must minimise this potential with
respect to the fields $\overline{H}$, $H$, $\overline{\theta}$, $\theta$,
and then set these minimisation relations equal to zero.

As these are set to zero, any linear combination of them is also zero, so
taking the following combinations and rearranging them, we have two
minimisation conditions
\begin{eqnarray}
\frac{\partial V}{\partial \overline{H}} -\frac{\partial V}{\partial H}
&\Rightarrow& \left\{\frac{1}{4} \left[ \frac{25}{9} g^2_F +g^2_{2R}
+\frac{3}{2} g^2_4 \right] (\overline{H} +H)^2 -\lambda^2_S (\overline{H}H
-M^2_H )\right\} (\overline{H} -H) \nonumber \\ & & \label{eq:minimise H}
-\frac{5}{6} g^2_F (\overline{\theta} +\theta)(\overline{\theta} -\theta)
(\overline{H} +H) = -m^2_{\overline{H}} \overline{H} +m^2_H H \\
\frac{\partial V}{\partial \overline{\theta}} -\frac{\partial V}{\partial
\theta} &\Rightarrow& \frac{1}{6}g^2_F \left[ 6(\overline{\theta}^2 -\theta^2 )
-5 (\overline{H}^2 -H^2)\right] (\overline{\theta} +\theta) \nonumber \\ & &
\label{eq:minimise theta}
-\lambda^2_{S'} (\overline{\theta} \theta -M^2_\theta) (\overline{\theta}
-\theta) = -m^2_{\overline{\theta}} \overline{\theta} +m^2_\theta \theta .
\end{eqnarray}

For $D$-flatness, it is necessary to set $\overline{H}^2 = H^2$ and
$\overline{\theta}^2 = \theta^2$, which results in $V_D = 0$. For $F_S$-flatness
and $F_{S'}$-flatness, it is necessary to set $\overline{H} H = M^2_H$ and
$\overline{\theta} \theta = M^2_\theta$, yielding zero valued F-terms. 
We wish to perturb away from these flatness conditions, so we impose
\begin{eqnarray}
\label{eq:perturb H}
\overline{H} = M_H -\overline{m} &;& H = M_H -m \\ \label{eq:perturb theta}
\overline{\theta} = M_\theta -\overline{m}' &;& \theta = M_\theta -m' ,
\end{eqnarray}
where $\overline{m}$, $m$, $\overline{m}'$, and $m'$ are all TeV scale masses.

Thus the two minimisation conditions, Eqs.~(\ref{eq:minimise H}) and
(\ref{eq:minimise theta}), become\footnote{After rearranging and taking
the leading order in the GUT scale masses $M_H$ and $M_\theta$.}
\begin{eqnarray}
\label{eq:min H 2}
g^2_H (m-\overline{m}) M_H -\frac{10}{3} g^2_F (m' -\overline{m}') M_\theta
&=& m^2_H -m^2_{\overline{H}} \\ \label{eq:min theta 2}
\frac{2}{3} g^2_F \left[ 6(m' -\overline{m}' )M_\theta -5(m-\overline{m})M_H
\right] &=& m^2_\theta - m^2_{\overline{\theta}}\; .
\end{eqnarray}

Now putting the small perturbations, Eqs.~(\ref{eq:perturb H}) and
(\ref{eq:perturb theta}), into Eqs.~(\ref{eq:D^2_H initial}) and
(\ref{eq:D^2_theta initial}) for the D-terms, we have, to leading order,
\begin{eqnarray}
\label{eq:D^2_H intermediate}
D^2_H &=& \frac{1}{4} (m-\overline{m})M_H \\ \label{eq:D^2_theta intermediate}
D^2_\theta &=& \frac{1}{3} \left[ 6(m' -\overline{m}')M_\theta -5(m-
\overline{m} )M_H \right] .
\end{eqnarray}

So, using Eqs.~(\ref{eq:min H 2}) and (\ref{eq:min theta 2}) in the above
Eqs.~(\ref{eq:D^2_H intermediate}) and (\ref{eq:D^2_theta intermediate}),
we have the following expressions for our D-terms for Model 1, as functions of the
soft SUSY breaking masses, GUT scale masses and gauge couplings
\begin{eqnarray}
\label{eq:D^2_H final}
D^2_H &=& \frac{1}{4g^2_{2R}+6g^2_4} \left[ m^2_H -m^2_{\overline{H}} +\frac{5}{6}
(m^2_\theta -m^2_{\overline{\theta}}) \right] \\ \label{eq:D^2_theta final}
D^2_\theta &=& \frac{m^2_\theta -m^2_{\overline{\theta}}}{2g^2_F} .
\end{eqnarray}
Note that Eq.~(\ref{eq:D^2_theta final}) was used in obtaining
Eq.~(\ref{eq:D^2_H final}). This is the form of the D-terms as used in
the updated version of SOFTSUSY \cite{Allanach:2001kg} to compute
the slepton mass data for our lepton flavour violating branching ratios
in Model 1, using Eqs.~(\ref{eq:m^2_QL}) to (\ref{eq:m^2_hd}).

For Model 2, the derivation is very similar, with the factor of $q_H$ being
the only difference, and so we obtain the form of $D^2_H$ below,
with $D^2_\theta$ being the same in both models.

\begin{eqnarray}
\label{eq:D^2_H final2}
D^2_H &=& \frac{(m^2_H -m^2_{\overline{H}})
+(m^2_\theta -m^2_{\overline{\theta}})}{4g^2_{2R}+6g^2_4} .
\end{eqnarray}

In both cases, we can see that the Pati-Salam limit is obtained when the
Froggatt-Nielsen scalar masses are degenerate, $m^2_\theta = m^2_{\overline{\theta}}$.
This result differs from \cite{King:2000vp} due to a different derivation
procedure.


\section{Operators for the two models}
\label{operatorsformodels}

In this appendix we give the operators which are responsible
for generating the Yukawa matrices of Models 1 and 2 analysed in this paper.

\subsection{Model 1}
\label{sec:model-1}


\begin{table}[tp]
  \centering
  \begin{tabular}{|c|c|c|c|}
    \hline  
    \multicolumn{2}{|c|}{Model 1} &
    \multicolumn{2}{c|}{Model 2} \\
    \hline
    $\delta$ & 0.22 & $\delta$ & 0.22 \\
    $\epsilon$ & 0.22 & $\epsilon$ & 0.22 \\
    \hline
    $a_{33}$ & 0.55 & $a_{33}$ & 0.55 \\
    \hline
    $a_{11}$ & -0.92 & $a_{11}^\prime$ & -0.92 \\
    $a_{12}$ & 0.33 & $a_{12}^\prime$ & 0.33 \\
    $a_{13}$ & 0.00 & $a_{13}$ & 0.00 \\
    $a_{21}$ & 1.67 & $a_{21}$ & 1.67 \\
    $a_{22}$ & 1.12 & $a_{22}$ & 1.12 \\
    $a_{23}$ & 0.89 & $a_{23}$ & 0.89 \\
    $a_{31}$ & -0.21 & $a_{31}$ & -0.21 \\
    $a_{32}$ & 2.08 & $a_{32}$ & 2.08 \\
    \hline
    $a^\prime_{12}$ & 0.77 & $a^{\prime\prime}_{12}$ & 0.77 \\
    $a^\prime_{13}$ & 0.53 & $a^{\prime\prime}_{13}$ & 0.53 \\
    $a^\prime_{22}$ & 0.66 & $a^\prime_{22}$ & 0.66 \\
    $a^\prime_{23}$ & 0.40 & $a^\prime_{23}$ & 0.40 \\
    $a^\prime_{32}$ & 1.80 & $a^\prime_{32}$ & 1.80 \\
    \hline
    $a^{\prime\prime}_{11}$ & 0.278 & $a^{\prime\prime\prime}_{11}$ & 0.278 \\
    $a^{\prime\prime}_{12}$ & 0.000 & $a^{\prime\prime\prime}_{12}$ & 0.000 \\
    $a^{\prime\prime}_{13}$ & 0.000 & $a^{\prime\prime\prime}_{13}$ & 0.000 \\
    \hline
    $A_{11}$ & 0.94 &     $A_{11}$ & 0.94  \\
    $A_{12}$ & 0.48 &    $A_{12}$ & 0.48 \\
    $A_{13}$ & 2.10 &    $A_{13}$ & 2.10 \\
    $A_{22}$ & 0.52 &    $A_{22}$ & 0.52 \\
    $A_{23}$ & 1.29 &    $A_{23}$ & 1.29 \\
    $A_{33}$ & 1.88 &    $A_{33}$ & 1.88 \\
    \hline
  \end{tabular}
  \caption{The $O(1)$ coefficients in Model 1 and Model 2. The values of $a_{13}$,
  $a^{\prime\prime}_{12}$ (in Model 1) and $a^{\prime\prime\prime}_{12}$ ( in Model 2)
will be varied, as discussed in the text.}
  \label{tab:a_ap_app_for_models}
\end{table}

This model is almost the model studied in
\cite{Blazek:2003wz,KP0307091}, but with an extra operator in the
$(1,2)$ and $(1,3)$ Yukawa matrix elements to allow a non-zero $Y^e_{12}$ 
and $Y^e_{13}$. The
operator texture is
{\footnotesize
\begin{equation}
  \label{eq:operator_texture1}
  \mathcal{O} 
  =
  \left[
    \begin{array}{ccc}
      ( a_{11} \mathcal{O}^{Fc}
      + a^{\prime\prime}_{11} \mathcal{O}''^{Ae} )\epsilon^5 &
      ( a_{12} \mathcal{O}^{Ee} + a^\prime_{12} \mathcal{O}'^{Cb}
      + a^{\prime\prime}_{12}O''^{Ec} ) \epsilon^3 &
      ( a_{13} \mathcal{O}^{Ec} + a^\prime_{13} \mathcal{O}'^{Cf}
      + a^{\prime\prime}_{13}\mathcal{O}''^{Ee} )\epsilon \\
      ( a_{21} \mathcal{O}^{Dc} ) \epsilon^4 &
      ( a_{22} \mathcal{O}^{Bc} + a_{22}^\prime \mathcal{O}'^{Ff} ) \epsilon^2 &
      ( a_{23} \mathcal{O}^{Ee} + a^\prime_{23} \mathcal{O}'^{Bc} )  \\
      ( a_{31} \mathcal{O}^{Fc} ) \epsilon^4 &
      ( a_{32} \mathcal{O}^{Ac} + a_{23}^\prime \mathcal{O}'^{Fe} ) \epsilon^2
      & a_{33}
    \end{array}
  \right]
\end{equation}
} 
where the operator nomenclature is defined in Appendix
\ref{sec:n=1-operators} and Appendix \ref{sec:n1-operators}.
This leads to the following Yukawa textures:
\begin{eqnarray}
  \label{eq:explicit_yu_model1}
  Y^{u}(M_X) &=&
  \left[
    \begin{array}{ccc}
    a^{\prime\prime}_{11} \sqrt{2} \delta^3 \epsilon^5 & 
    a^\prime_{12} \sqrt{2} \delta^2\epsilon^3 &  
    a^\prime_{13}\frac{2}{\sqrt{5}} \delta^2\epsilon  \\
    0 & 
    a^\prime_{22} \frac{8}{5 \sqrt{5}}\delta^2\epsilon^2 
    & 0 \\
    0 
    & a^\prime_{32} \frac{8}{5} \delta^2 \epsilon^2 
    & a_{33}
    \end{array}
  \right]
  \\
  \label{eq:explicit_yd_model1}
  Y^{d}(M_X) &=&
  \left[
    \begin{array}{ccc}
      a_{11} \frac{8}{5}\delta \epsilon^5 & -a^\prime_{12} \sqrt{2}
      \delta^2\epsilon^3 & a^\prime_{13} \frac{4}{\sqrt{5}} \delta^2\epsilon\\ a_{21}
      \frac{2}{\sqrt{5}}\delta \epsilon^4 & (a_{22} \sqrt{\frac{2}{5}}
      \delta + a^\prime_{22} \frac{16}{5\sqrt{5}} \delta^2)\epsilon^2
      & a^\prime_{23} \sqrt{\frac{2}{5}} \delta^2 \\ a_{31}
      \frac{8}{5}\delta\epsilon^4 & a_{32} \sqrt{2} \delta \epsilon^2
      & a_{33}
    \end{array}
  \right]
  \\
  \label{eq:explicit_ye_model1}
  Y^{e}(M_X) &=&
  \left[
    \begin{array}{ccc}
    a_{11} \frac{6}{5}\delta \epsilon^5 & a_{12}^{\prime\prime} 2\delta^2 \epsilon^3 &
    a_{13} 2 \delta^2 \epsilon \\ a_{21} \frac{4}{\sqrt{5}} \delta\epsilon^4 & ( -a_{22}
    3\sqrt{\frac{2}{5}} \sqrt{\frac{2}{5}} + a^\prime_{22} \delta
    \frac{12}{5\sqrt{5}} ) \delta\epsilon^2 & -a^\prime_{23}
    \sqrt{\frac{2}{5}} \delta^2 \\ -a_{31} \frac{6}{5}
    \delta\epsilon^4 & a_{32} \sqrt{2} \delta\epsilon^2 & a_{33}
  \end{array}
\right]
  \\
  \label{eq:explicit_yn_model1}
  Y^{\nu}(M_X) &=&
  \left[
    \begin{array}{ccc}
      a^{\prime\prime}_{11} \sqrt{2}\delta^3\epsilon^5 & 
      a_{12} 2 \delta\epsilon^3 & 
      a^{\prime\prime}_{13} 2\delta^3\epsilon \\
      0 & 
      a^\prime_{22} \frac{6}{5\sqrt{5}}\delta^2\epsilon^2 & 
      a_{23} 2 \delta \\
      0 & 
      a^\prime_{32} \frac{6}{5} \delta^2\epsilon^2 & 
      a_{33}
    \end{array}
  \right]
\end{eqnarray}

For both models we define $\epsilon$ and $\delta$ as:
\begin{equation}
  \label{eq:def_eps_delta} \epsilon = \left( \frac{\langle\theta\rangle}{M_X}
\right) \;\;\;\; ; \;\;\;\; \delta = \left( \frac{\langle H\rangle\langle\overline
H\rangle}{M^2_X} \right).
\end{equation}
We take $\delta = \epsilon = 0.22$.

The values of the arbritary couplings are laid out in
Table~\ref{tab:a_ap_app_for_models}. This gives numerical values for
the Yukawa elements which can be used in either model, with the relevant
values of $Y^e_{12}$ and $Y^e_{13}$ inserted instead of the texture zeros:
\begin{eqnarray}
  \label{eq:numerical_yu_model1A}
  Y^{u}(M_X) &=&
  \left[
    \begin{array}{lll}
      2.159\times10^{-06}  &  5.606\times10^{-04}  &  5.090\times10^{-03} \\
      0.000    &    1.105\times10^{-03}  &   0.000    \\
      0.000    &    6.733\times10^{-03}  &  5.841\times10^{-01}    \\
    \end{array}
  \right]
  \\
  \label{eq:numerical_yd_model1A}
  Y^{d}(M_X) &=&
  \left[
    \begin{array}{lll}
    \!\!\!\!-1.661\times10^{-04} & \!\!\!\! -5.606\times10^{-04} &   1.018\times10^{-02} \\
     7.683\times10^{-04}  & \!\!\!\!-5.343\times10^{-03}  &  1.216\times10^{-02}\\
    \!\!\!\!-1.769\times10^{-04}   & 3.133\times10^{-02}   & 3.933\times10^{-01}\\
    \end{array}
  \right]
  \\
  \label{eq:numerical_ye_model1A}
  Y^{e}(M_X) &=&
  \left[
    \begin{array}{lll}
    \!\!\!\!-1.246\times10^{-04} &    0.000   &      0.000   \\
     1.537\times10^{-03}  &  2.432\times10^{-02}&   \!\!\!\!-3.649\times10^{-02}\\
    \!\!\!\!-1.327\times10^{-04}   & 3.133\times10^{-02} &   5.469\times10^{-01}\\
  \end{array}
\right]
  \\
  \label{eq:numerical_yn_model1A}
  Y^{\nu}(M_X) &=&
  \left[
    \begin{array}{lll}
     2.159\times10^{-06}&    1.525\times10^{-03}&    0.000\\
      0.000     &   8.290\times10^{-04} &   3.923\times10^{-01}  \\  
      0.000      &  5.050\times10^{-03}  &  5.469\times10^{-01} \\
    \end{array}
  \right]
\end{eqnarray}

The RH Majorana neutrino mass matrix is:
\begin{equation}
  \label{eq:21}
  \frac{M_{RR}(M_X)}{M_{33}} =
  \left[
    \begin{array}{ccc}
      A_{11}\delta\epsilon^8 & A_{12}\delta\epsilon^6 &
      A_{13}\delta\epsilon^4 \\ A_{12}\delta\epsilon^6 &
      A_{22}\delta\epsilon^4 & A_{23}\delta\epsilon^2 \\
      A_{13}\delta\epsilon^4 & A_{23}\delta\epsilon^2 & A_{33}
    \end{array}
  \right]
\end{equation}

The numerical values for the Majorana mass matrix are
\begin{equation}
  \label{eq:numerical_21A}
  \frac{M_{RR}(M_X)}{M_{33}} =
  \left[
    \begin{array}{lll}
      3.508\times10^{8}  &       3.686\times10^{9}   &     3.345\times10^{11}\\
      3.686\times10^{9}   &     8.313\times10^{10}&    5.886\times10^{12}\\
     3.345\times10^{11}&    5.886\times10^{12} &   5.795\times10^{14}\\
    \end{array}
  \right]
\end{equation}


\subsection{Model 2}
\label{sec:model-2}


The operator texture for Model 2 is:
{\footnotesize
\begin{equation}
  \label{eq:operator_texture2}
  \mathcal{O} 
  =
  \left[
    \begin{array}{ccc}
      ( a_{11}^\prime \mathcal{O}'^{Fc}
      + a^{\prime\prime\prime}_{11} \mathcal{O}'''^{Ae} )\epsilon^4 &
      ( a_{12}^\prime \mathcal{O}'^{Ee} + a^{\prime\prime}_{12} \mathcal{O}''^{Cb}
      + a^{\prime\prime\prime}_{12}O'''^{Ec} ) \epsilon^2 &
      ( a_{13} \mathcal{O}^{Ec} + a^{\prime\prime}_{13} \mathcal{O}''^{Cf}
      + a^{\prime\prime\prime}_{13}\mathcal{O}'''^{Ee} )  \\
      ( a_{21} \mathcal{O}^{Dc} ) \epsilon^4 &
      ( a_{22} \mathcal{O}^{Bc} + a_{22}^\prime \mathcal{O}'^{Ff} ) \epsilon^2 &
      ( a_{23} \mathcal{O}^{Ee} + a^\prime_{23} \mathcal{O}'^{Bc} )  \\
      ( a_{31} \mathcal{O}^{Fc} ) \epsilon^4 &
      ( a_{32} \mathcal{O}^{Ac} + a_{23}^\prime \mathcal{O}'^{Fe} ) \epsilon^2
      & a_{33}
    \end{array}
  \right]
\end{equation}
}

The operator nomenclature is defined in Appendix \ref{sec:n=1-operators} and
Appendix \ref{sec:n1-operators}. The new operator setup leads to the following
Yukawa textures:
\begin{eqnarray}
  \label{eq:explicit_yu_model2}
  Y^{u}(M_X) &=&
  \left[
    \begin{array}{ccc}
    a^{\prime\prime\prime}_{11} \sqrt{2} \delta^4 \epsilon^4 & 
    a^{\prime\prime}_{12} \sqrt{2} \delta^3\epsilon^2 &  
    a^{\prime\prime}_{13}\frac{2}{\sqrt{5}} \delta^3  \\
    0 & 
    a^\prime_{22} \frac{8}{5 \sqrt{5}}\delta^2\epsilon^2 
    & 0 \\
    0 
    & a^\prime_{32} \frac{8}{5} \delta^2 \epsilon^2 
    & a_{33}
    \end{array}
  \right]
  \\
  \label{eq:explicit_yd_model2}
  Y^{d}(M_X) &=&
  \left[
    \begin{array}{ccc}
      a^\prime_{11} \frac{8}{5}\delta^2 \epsilon^4 & 
      -a^{\prime\prime}_{12} \sqrt{2} \delta^3\epsilon^2 &
      a^{\prime\prime}_{13} \frac{4}{\sqrt{5}} \delta^3\\ 
      a_{21} \frac{2}{\sqrt{5}}\delta \epsilon^4 & (a_{22} \sqrt{\frac{2}{5}}
      \delta + a^\prime_{22} \frac{16}{5\sqrt{5}} \delta^2)\epsilon^2
      & a^\prime_{23} \sqrt{\frac{2}{5}} \delta^2 \\ a_{31}
      \frac{8}{5}\delta\epsilon^4 & a_{32} \sqrt{2} \delta \epsilon^2
      & a_{33}
    \end{array}
  \right]
  \\
  \label{eq:explicit_ye_model2}
  Y^{e}(M_X) &=&
  \left[
    \begin{array}{ccc}
    a_{11}^\prime \frac{6}{5}\delta^2 \epsilon^4 & a_{12}^{\prime\prime\prime} 2 \delta^3
    \epsilon^2 & a_{13} 2 \delta^3 \\ a_{21}
    \frac{4}{\sqrt{5}} \delta\epsilon^4 & ( -a_{22}
    3\sqrt{\frac{2}{5}} \sqrt{\frac{2}{5}} + a^\prime_{22} \delta
    \frac{12}{5\sqrt{5}} ) \delta\epsilon^2 & -a^\prime_{23}
    \sqrt{\frac{2}{5}} \delta^2 \\ -a_{31} \frac{6}{5}
    \delta\epsilon^4 & a_{32} \sqrt{2} \delta\epsilon^2 & a_{33}
  \end{array}
\right]
  \\
  \label{eq:explicit_yn_model2}
  Y^{\nu}(M_X) &=&
  \left[
    \begin{array}{ccc}
      a^{\prime\prime\prime}_{11} \sqrt{2}\delta^4\epsilon^4 & 
      a^\prime_{12} 2 \delta^2\epsilon^2 & 
      a^{\prime\prime\prime}_{13} 2\delta^4 \\
      0 & 
      a^\prime_{22} \frac{6}{5\sqrt{5}}\delta^2\epsilon^2 & 
      a_{23} 2 \delta \\
      0 & 
      a^\prime_{32} \frac{6}{5} \delta^2\epsilon^2 & 
      a_{33}
    \end{array}
  \right]
\end{eqnarray}

The RH Majorana neutrino mass matrix is the same as in Model 1:
\begin{equation}
\label{eq:4}
\frac{M_{RR}(M_X)}{M_{33}} =
  \left[
    \begin{array}{ccc}
      A_{11}\delta\epsilon^8 & A_{12}\delta\epsilon^6 &
      A_{13}\delta\epsilon^4 \\ A_{12}\delta\epsilon^6 &
      A_{22}\delta\epsilon^4 & A_{23}\delta\epsilon^2 \\
      A_{13}\delta\epsilon^4 & A_{23}\delta\epsilon^2 & A_{33}
    \end{array}
  \right]
\end{equation}


\section{$n=1$ operators}
\label{sec:n=1-operators}


\begin{table}[htbp]
  \centering
  \mbox
  {
  \begin{tabular}{|c|c||c|c|c|c|}
    \hline Operator Name & Operator Name in
    \cite{King:OperatorAnalysis} & $Q\overline{U}h_2$ &
    $Q\overline{D}h_1$ & $L \overline{E}h_1$ & $L \overline{N}h_2$ \\
    \hline $O^{Aa}$ & $O^A$ & $1$ & $1$ & $1$ & $1$ \\ $O^{Ab}$ &
    $O^B$ & $1$ & $-1$ & $-1$ & $1$ \\ $O^{Ac}$ & $O^M$ & $0$ &
    $\sqrt{2}$ & $\sqrt{2}$ & $0$ \\ $O^{Ad}$ & $O^T$ &
    $\frac{2\sqrt{2}}{5}$ & $\frac{\sqrt{2}}{5}$ &
    $\frac{\sqrt{2}}{5}$ & $\frac{2\sqrt{2}}{5}$ \\ $O^{Ae}$ & $O^V$ &
    $\sqrt{2}$ & $0$ & $0$ & $\sqrt{2}$ \\ $O^{Af}$ & $O^U$ &
    $\frac{\sqrt{2}}{5}$ & $\frac{2\sqrt{2}}{5}$ &
    $\frac{2\sqrt{2}}{5}$ & $\frac{\sqrt{2}}{5}$ \\ \rule{0mm}{4mm}
    $O^{Ba}$ & $O^C$ & $\frac{1}{\sqrt{5}}$ & $\frac{1}{\sqrt{5}}$ &
    $\frac{-3}{\sqrt{5}}$ & $\frac{-3}{\sqrt{5}}$ \\ \rule{0mm}{4mm}
    $O^{Bb}$ & $O^D$ & $\frac{1}{\sqrt{5}}$ & $\frac{-1}{\sqrt{5}}$ &
    $\frac{-3}{\sqrt{5}}$ & $\frac{3}{\sqrt{5}}$ \\ $O^{Bc}$ & $O^W$ &
    $0$ & $\sqrt{\frac{2}{5}}$ & $-3\sqrt{\frac{2}{5}}$ & $0$ \\
    $O^{Bd}$ & $O^X$ & $\frac{2\sqrt{2}}{5}$ & $\frac{\sqrt{2}}{5}$ &
    $\frac{-3\sqrt{2}}{5}$ & $\frac{-6\sqrt{2}}{5}$ \\ $O^{Be}$ &
    $O^Z$ & $\sqrt{\frac{2}{5}}$ & $0$ & $0$ & $-3\sqrt{\frac{2}{5}}$
    \\ $O^{Bf}$ & $O^Y$ & $\frac{\sqrt{2}}{5}$ & $\frac{2\sqrt{2}}{5}$
    & $\frac{-6\sqrt{2}}{5}$ & $\frac{-3\sqrt{2}}{5}$ \\ $O^{Ca}$ &
    $O^a$ & $\sqrt{2}$ & $\sqrt{2}$ & $0$ & $0$ \\ $O^{Cb}$ & $O^F$ &
    $\sqrt{2}$ & $-\sqrt{2}$ & $0$ & $0$ \\ $O^{Cc}$ & $O^E$ & $0$ &
    $2$ & $0$ & $0$ \\ $O^{Cd}$ & $O^b$ & $\frac{4}{\sqrt{5}}$ &
    $\frac{2}{\sqrt{5}}$ & $0$ & $0$ \\ $O^{Ce}$ & $O^N$ & $2$ & $0$ &
    $0$ & $0$ \\ $O^{Cf}$ & $O^c$ & $\frac{2}{\sqrt{5}}$ &
    $\frac{4}{\sqrt{5}}$ & $0$ & $0$ \\ $O^{Da}$ & $O^d$ &
    $\sqrt{\frac{2}{5}}$ & $\sqrt{\frac{2}{5}}$ &
    $2\sqrt{\frac{2}{5}}$ & $2\sqrt{\frac{2}{5}}$ \\ $O^{Db}$ & $O^e$
    & $\sqrt{\frac{2}{5}}$ & $-\sqrt{\frac{2}{5}}$ &
    $-2\sqrt{\frac{2}{5}}$ & $2\sqrt{\frac{2}{5}}$ \\ $O^{Dc}$ & $O^G$
    & $0$ & $\frac{2}{\sqrt{5}}$ & $\frac{4}{\sqrt{5}}$ & $0$ \\
    \rule{0mm}{4mm} $O^{Dd}$ & $O^H$ & $\frac{4}{5}$ & $\frac{2}{5}$ &
    $\frac{4}{5}$ & $\frac{8}{5}$ \\ \rule{0mm}{4mm} $O^{De}$ & $O^O$
    & $\frac{2}{\sqrt{5}}$ & $0$ & $0$ & $\frac{4}{\sqrt{5}}$ \\
    $O^{Df}$ & $O^f$ & $\frac{2}{5}$ & $\frac{4}{5}$ & $\frac{8}{5}$ &
    $\frac{4}{5}$ \\ $O^{Ea}$ & $O^g$ & $0$ & $0$ & $\sqrt{2}$ &
    $\sqrt{2}$ \\ $O^{Eb}$ & $O^h$ & $0$ & $0$ & $-\sqrt{2}$ &
    $\sqrt{2}$ \\ $O^{Ec}$ & $O^i$ & $0$ & $0$ & $2$ & $0$ \\ $O^{Ed}$
    & $O^j$ & $0$ & $0$ & $\frac{2}{\sqrt{5}}$ & $\frac{4}{\sqrt{5}}$
    \\ $O^{Ee}$ & $O^I$ & $0$ & $0$ & $0$ & $2$ \\ $O^{Ef}$ & $O^J$ &
    $0$ & $0$ & $\frac{4}{\sqrt{5}}$ & $\frac{2}{\sqrt{5}}$ \\
    $O^{Fa}$ & $O^P$ & $\frac{4\sqrt{2}}{5}$ & $\frac{4\sqrt{2}}{5}$ &
    $\frac{3\sqrt{2}}{5}$ & $\frac{3\sqrt{2}}{5}$ \\ $O^{Fb}$ & $O^Q$
    & $\frac{4\sqrt{2}}{5}$ & $\frac{-4\sqrt{2}}{5}$ &
    $\frac{-3\sqrt{2}}{5}$ & $\frac{3\sqrt{2}}{5}$ \\ \rule{0mm}{4mm}
    $O^{Fc}$ & $O^R$ & $0$ & $\frac{8}{5}$ & $\frac{6}{5}$ & $0$ \\
    \rule{0mm}{4mm} $O^{Fd}$ & $O^L$ & $\frac{16}{5\sqrt{5}}$ &
    $\frac{8}{5\sqrt{5}}$ & $\frac{6}{5\sqrt{5}}$ &
    $\frac{12}{5\sqrt{5}}$ \\ $O^{Fe}$ & $O^K$ & $\frac{8}{5}$ & $0$ &
    $0$ & $\frac{6}{5}$ \\ $O^{Ff}$ & $O^S$ & $\frac{8}{5\sqrt{5}}$ &
    $\frac{16}{5\sqrt{5}}$ & $\frac{12}{5\sqrt{5}}$ &
    $\frac{6}{5\sqrt{5}}$ \\ \hline
  \end{tabular}
  }
  \caption{Operator names, CGCs and names in \cite{King:OperatorAnalysis}}
  \label{tab:operator_names}
\end{table}

The $n=1$ Dirac operators are the complete set of all operators that can be
constructed from the quintilinear $F\overline{F}h\overline{H}H$ by all possible
group theoretical contractions of the indicies in
\begin{equation}
  \label{eq:fields_tensor_app}
  \mathcal{O}^{\alpha \rho y w}_{\beta \gamma x z} =
  F^{\alpha a}\overline{F}_{\beta x} h^y_a \overline{H}_{\gamma z} H^{\rho w}
\end{equation}

We define some \SU{4} invariant tensors $C$ and some \SU{2} invariant
tensors $R$ as follows\footnote{The subscript denotes the dimension of
the representation they can create from multiplying $\mathbf{4}$ or
$\overline{\mathbf{4}}$ with $\mathbf{4}$ or
$\overline{\mathbf{4}}$. For example
$(C_{15})^{\beta\gamma}_{\alpha\rho}\overline{\mathbf{4}}_\gamma
\mathbf{4}^\rho = \mathbf{15}^\beta_\alpha$ .}:
\begin{eqnarray}
  \nonumber \left(C_1\right)^\alpha_\beta &=& \delta^\alpha_\beta \\
  \nonumber \left(C_6\right)^{\rho\gamma}_{\alpha\beta} &=&
  \epsilon_{\alpha\beta\omega\chi}^{\rho\gamma\omega\chi} \\ \nonumber
  \left(C_{10}\right)^{\alpha\beta}_{\rho\gamma} &=&
  \delta^\alpha_\rho \delta^\beta_\gamma + \delta^\alpha_\gamma
  \delta^\beta_\rho \\ \nonumber
  \left(C_{15}\right)^{\beta\gamma}_{\alpha\rho} &=& \delta^\beta_\rho
  \delta^\gamma_\alpha -\frac{1}{4} \delta^\beta_\alpha
  \delta^\gamma_\rho \\ \nonumber \left(R_1\right)^x_y &=& \delta^x_y
  \\
  \label{eq:tensors}
  \left(R_3\right)^{wx}_{yz} &=& 
  \delta^x_y \delta^w_z - \frac{1}{2} \delta^x_z \delta^w_y
\end{eqnarray}

Then the six independent \SU{4} structures are:
\begin{eqnarray}
  \nonumber
  \mathrm{A}. &
  \left(C_1\right)^\beta_\alpha \left(C_1\right)^\gamma_\rho 
  &=\;\;
  \delta^\beta_\alpha \delta^\gamma_\rho
  \\
  \nonumber
  \mathrm{B}. &
  \left(C_{15}\right)^{\beta\chi}_{\alpha\sigma}
  \left(C_{15}\right)^{\gamma\sigma}_{\rho\chi}
  &=\;\;
  \delta^\beta_\rho \delta^\gamma_\alpha - 
  \frac{1}{4} \delta^\beta_\alpha \delta^\gamma_\rho
  \\
  \nonumber
  \mathrm{C}. &
  \left(C_6\right)^{\omega\chi}_{\alpha\rho}
  \left(C_6\right)^{\beta\gamma}_{\omega\chi} 
  &=\;\;
  8(\delta^\beta_\alpha \delta^\gamma_\alpha 
  - \delta^\gamma_\alpha \delta^\beta_\rho )
  \\
  \nonumber
  \mathrm{D}. &
  \left(C_{10}\right)^{\omega\chi}_{\alpha\rho}
  \left(C_{10}\right)^{\beta\gamma}_{\omega\chi} 
  &=\;\;
  2(\delta^\beta_\alpha \delta^\gamma_\rho
  + \delta^\gamma_\alpha \delta^\beta_\rho )
  \\
  \nonumber
  \mathrm{E}. &
  \left(C_1\right)^\beta_\rho
  \left(C_1\right)^\gamma_\alpha 
  &=\;\;
  \delta^\beta_\alpha \delta^\gamma_\alpha
  \\
  \label{eq:su4_structures}
  \mathrm{F}. &
  \left(C_{15}\right)^{\gamma\chi}_{\alpha\sigma}
  \left(C_{15}\right)^{\beta\sigma}_{\rho\chi}
  &=\;\;
  \delta^\gamma_\rho \delta^\alpha_\beta
  -\frac{1}{4}\delta^\gamma_\alpha \delta^\beta_\rho
\end{eqnarray}

And the six \SU{2} structures are:
\begin{eqnarray}
  \nonumber
  \mathrm{a}. &
  \left(R_1\right)^z_w
  \left(R_1\right)^x_y 
  &= \;\;
  \delta^z_w \delta^x_y
  \\
  \nonumber
  \mathrm{b}. &
  \left(R_3\right)^{zq}_{wr}
  \left(R_3\right)^{xr}_{yq} 
  &= \;\;
  \delta^x_w \delta^z_y - \frac{1}{2}\delta^x_y\delta^z_w
  \\
  \nonumber
  \mathrm{c}. &
  \epsilon^{xz}\epsilon_{yw} 
  &= \;\;
  \epsilon^{xz}\epsilon_{yw}
  \\
  \nonumber
  \mathrm{d}. &
  \epsilon_{ws}\epsilon^{xt}
  \left(R_3\right)^{sq}_{yr}
  \left(R_3\right)^{zr}_{tq} 
  &= \;\;
  \delta^x_w\delta^z_y - \frac{1}{2}\epsilon_{wy}\epsilon^{xz}
  \\
  \nonumber 
  \mathrm{e}. &
  \left(R_1\right)^z_y 
  \left(R_1\right)^x_w  
  &= \;\;
  \delta^z_y \delta^x_w
  \\
  \label{eq:su2_structurres}
  \mathrm{f}. &
  \left(R_3\right)^{zq}_{yr}
  \left(R_3\right)^{xr}_{wq} 
  &= \;\;
  \delta^x_y\delta^z_w - \frac{1}{2}\delta^x_w\delta^z_y
\end{eqnarray}

All possible $n=1$ operators were then named $O^A ... O^Z O^a...O^j$
in \cite{King:OperatorAnalysis}. We rename them here in a manner
consistent with the $n>1$ operators $O^{(n')}$, so that the names are
$O^{\Pi\pi}$ where $\Pi$ is the \SU{4} structure and $\pi$ is the
SU{2} structure.  See Table \ref{tab:operator_names} for the
translation into the names of Ref.\cite{King:OperatorAnalysis} and the
CGCs.

All of these operators are operators for the case without a \UI{}
family symmetry. In the case when there is, we follow the
prescription
\begin{equation}
  \label{eq:u1isation_of_operators}
  \mathcal{O}_{IJ}\rightarrow \mathcal{O}_{IJ}
  \left(\frac{\theta}{M_X}\right) ^{p_{_{IJ}}} .
\end{equation}

Where $p_{IJ} = |X_{\mathcal{O}_{IJ}}|$ is the modulus of the charge
of the operator. If the charge of the operator is negative, then the
field $\theta$ should be replaced by the field
$\overline{\theta}$. The prescription makes the operator chargeless
under the $\UI_F$ while simultaneously not changing the dimension.


\section{$n>1$ operators}
\label{sec:n1-operators}


In the case that $n > 1$, there will be more indicies to contract,
which allows more representations, and hence more Clebsch
coefficients. To generalise the notation, it is necessary only to
construct the new tensors which create the new structures. However, it
will always be possible to contract the new indicies between the $H$
and $\overline{H}$ fields to create a singlet $H\overline{H}$ which
has a Clebsch of 1 in each sector $u,d,e,\nu$. In this case, the first
structures are the same as the old structures, but with extra $\delta$
symbols which construct the $H\overline{H}$ singlet.

Thus taking an $n=2$ operator, say $\mathcal{O}'^{Fb}$, which forms a
representation that could have been attained by a $n=1$ operator, the
Clebsch coefficients are the same. This is what we mean by
$\mathcal{O}^{n\prime\;\Pi\pi}$, as we have only used $n>1$
coefficients which are in the subset that have $n=1$ analogues.


\begin{thebibliography}{99}
\bibitem{Borzumati:1986qx}
F.~Borzumati and A.~Masiero,
Phys.\ Rev.\ Lett.\  {\bf 57} (1986) 961.
\bibitem{Gabbiani:1996hi}
F.~Gabbiani, E.~Gabrielli, A.~Masiero and L.~Silvestrini,
Nucl.\ Phys.\ B {\bf 477} (1996) 321
[arXiv:hep-ph/9604387].
\bibitem{Chung:2003fi}
D.~J.~H.~Chung, L.~L.~Everett, G.~L.~Kane, S.~F.~King, J.~Lykken and L.~T.~Wang,
Phys.\ Rept.\  {\bf 407} (2005) 1
[arXiv:hep-ph/0312378].
\bibitem{GoPe}
G.~L.~Fogli, E.~Lisi, A.~Marrone, D.~Montanino, A.~Palazzo and A.~M.~Rotunno,
arXiv:hep-ph/0212127;
P.~C.~de~Holanda and A.~Y.~Smirnov,
arXiv:hep-ph/0212270;
V.~Barger and D.~Marfatia,
Phys.\ Lett.\ B {\bf 555} (2003) 144
[arXiv:hep-ph/0212126];
A.~Bandyopadhyay, S.~Choubey, R.~Gandhi, S.~Goswami and D.~P.~Roy,
arXiv:hep-ph/0212146.
M.~Maltoni, T.~Schwetz and J.~W.~Valle,
arXiv:hep-ph/0212129.
\bibitem{SKamiokandeColl}
               Y. Fukuda {\it et al.}, Super-Kamiokande Collaboration,
               Phys. Lett. {\bf B433}, 9 (1998);
        {\it ibid.}\ Phys. Lett. {\bf B436}, 33 (1998);
        {\it ibid.}\ Phys. Rev. Lett. {\bf 81}, 1562 (1998).
\bibitem{King:2003jb}
  For a review see:
  S.~F.~King,
  Rept.\ Prog.\ Phys.\  {\bf 67}, 107 (2004)
  [arXiv:hep-ph/0310204].
\bibitem{Hisano:1995cp}
J.~Hisano, T.~Moroi, K.~Tobe and M.~Yamaguchi,
Phys.\ Rev.\ D {\bf 53} (1996) 2442
[arXiv:hep-ph/9510309].
\bibitem{King:1998nv}
S.~F.~King and M.~Oliveira,
Phys.\ Rev.\ D {\bf 60} (1999) 035003
[arXiv:hep-ph/9804283].
\bibitem{Blazek:2002wq}
T.~Blazek and S.~F.~King,
arXiv:hep-ph/0211368.
\bibitem{huge}
S.~Davidson and A.~Ibarra,
JHEP {\bf 0109} (2001) 013;
J.~Hisano and D.~Nomura,
Phys.\ Rev.\ D {\bf 59} (1999) 116005;
J.~Hisano,
arXiv:hep-ph/0204100;
J.~A.~Casas and A.~Ibarra,
Nucl.\ Phys.\ B {\bf 618} (2001) 171;
W.~Buchm\"uller, D.~Delepine and F.~Vissani,
Phys.\ Lett.\ B {\bf 459} (1999) 171;
M.~E.~Gomez, G.~K.~Leontaris, S.~Lola and J.~D.~Vergados,
Phys.\ Rev.\ D {\bf 59} (1999) 116009;
J.~R.~Ellis, M.~E.~Gomez, G.~K.~Leontaris, S.~Lola and D.~V.~Nanopoulos,
Eur.\ Phys.\ J.\ C {\bf 14} (2000) 319;
W.~Buchm\"uller, D.~Delepine and L.~T.~Handoko,
Nucl.\ Phys.\ B {\bf 576} (2000) 445;
D. Carvalho, J. Ellis, M. Gomez and S. Lola,
Phys.\ Lett.\ B {\bf 515} (2001) 323;
F.~Deppisch, H.~Pas, A.~Redelbach, R.~Ruckl and Y.~Shimizu,
arXiv:hep-ph/0206122;
J.~Sato and K.~Tobe,
Phys.\ Rev.\ D {\bf 63} (2001) 116010;
J.~Hisano and K.~Tobe,
Phys.\ Lett.\ B {\bf 510} (2001) 197;
J.~R.~Ellis, J.~Hisano, M.~Raidal and Y.~Shimizu,
Phys.\ Lett.\ B {\bf 528} (2002) 86,
arXiv:hep-ph/0111324;
J.~R.~Ellis, J.~Hisano, S.~Lola and M.~Raidal,
Nucl.\ Phys.\ B {\bf 621} (2002) 208,
arXiv:hep-ph/0109125;
J.~Hisano, T.~Moroi, K.~Tobe and M.~Yamaguchi,
Phys.\ Lett.\ B {391} (1997) 341; 
[Erratum - {\it ibid.} {\bf 397}, 357 (1997)];
J.~Hisano, D.~Nomura, Y.~Okada, Y.~Shimizu and M.~Tanaka,
Phys.\ Rev.\ D {\bf 58} (1998) 116010;
J.~Hisano, D.~Nomura and T.~Yanagida,
Phys.\ Lett.\ B {\bf 437} (1998) 351;
S.~Lavignac, I.~Masina and C.~A.~Savoy,
Phys.\ Lett.\ B {\bf 520} (2001) 269
[arXiv:hep-ph/0106245];
S.~Lavignac, I.~Masina and C.~A.~Savoy,
Nucl.\ Phys.\ B {\bf 633} (2002) 139
[arXiv:hep-ph/0202086];
I.~Masina and C.~A.~Savoy,
arXiv:hep-ph/0211283.
S.~Pascoli, S.~T.~Petcov and C.~E.~Yaguna,
Phys.\ Lett.\ B {\bf 564} (2003) 241
[arXiv:hep-ph/0301095].
S.~Pascoli, S.~T.~Petcov and W.~Rodejohann,
arXiv:hep-ph/0302054.
\bibitem{Chankowski:2005jh}
  P.~H.~Chankowski, O.~Lebedev and S.~Pokorski,
  Nucl.\ Phys.\ B {\bf 717} (2005) 190
  [arXiv:hep-ph/0502076].
\bibitem{Abel:2001ur}
S.~A.~Abel and G.~Servant,
Nucl.\ Phys.\ B {\bf 611} (2001) 43
[arXiv:hep-ph/0105262].
\bibitem{Ross:2002mr}
G.~G.~Ross and O.~Vives,
Phys.\ Rev.\ D {\bf 67} (2003) 095013
[arXiv:hep-ph/0211279].
\bibitem{KP0307091}
S.~F.~King, I.~N.~R.~Peddie,
Nucl.\ Phys.\ B {\bf 678} (2004) 339
[arXiv:hep-ph/0307091].
\bibitem{P_Matrix:introduction}
S.~A.~Abel, B.~C.~Allanach, F.~Quevedo, L.~Ibanez and M.~Klein,
JHEP {\bf 0012} (2000) 026
[arXiv:hep-ph/0005260].
B.~C.~Allanach, D.~Grellscheid and F.~Quevedo,
JHEP {\bf 0205} (2002) 048
[arXiv:hep-ph/0111057].
\bibitem{Brignole:1997dp}
A.~Brignole, L.~E.~Ibanez and C.~Munoz,
[arXiv:hep-ph/9707209].
\bibitem{Ibanez:1998rf}
L.~E.~Ibanez, C.~Munoz and S.~Rigolin,
Nucl.\ Phys.\ B {\bf 553} (1999) 43
[arXiv:hep-ph/9812397].
\bibitem{King:2003xn}
S.~F.~King and I.~N.~R.~Peddie,
J.\ Korean Phys.\ Soc.\  {\bf 45} (2004) S443
[arXiv:hep-ph/0312235].
\bibitem{Everett:2002pm}
L.~L.~Everett, G.~L.~Kane, S.~F.~King, S.~Rigolin and L.~T.~Wang,
Phys.\ Lett.\ B {\bf 531} (2002) 263
[arXiv:hep-ph/0202100].
\bibitem{Blazek:2003wz}
T.~Blazek, S.~F.~King and J.~K.~Parry,
arXiv:hep-ph/0303192.
\bibitem{King:OperatorAnalysis}
S.~F.~King,
Phys.\ Lett.\ B {\bf 325} (1994) 129
[Erratum-ibid.\ B {\bf 325} (1994) 538];
B.~C.~Allanach and S.~F.~King,
Nucl.\ Phys.\ B {\bf 456} (1995) 57
[arXiv:hep-ph/9502219];
B.~C.~Allanach and S.~F.~King,
Nucl.\ Phys.\ B {\bf 459} (1996) 75
[arXiv:hep-ph/9509205];
B.~C.~Allanach, S.~F.~King, G.~K.~Leontaris and S.~Lola,
Phys.\ Rev.\ D {\bf 56} (1997) 2632
[arXiv:hep-ph/9610517];
S.~F.~King and M.~Oliveira,
Phys.\ Rev.\ D {\bf 63} (2001) 095004
[arXiv:hep-ph/0009287]
\bibitem{Froggatt:1978nt}
C.~D.~Froggatt and H.~B.~Nielsen,
Nucl.\ Phys.\ B {\bf 147} (1979) 277.
\bibitem{King:2000vp}
S.~F.~King and M.~Oliveira,
Phys.\ Rev.\ D {\bf 63} (2001) 015010
[arXiv:hep-ph/0008183].
\bibitem{Kane:2005va}
  G.~L.~Kane, S.~F.~King, I.~N.~R.~Peddie and L.~Velasco-Sevilla,
  arXiv:hep-ph/0504038.
\bibitem{Abel:2001cv}
S.~Abel, S.~Khalil and O.~Lebedev,
Phys.\ Rev.\ Lett.\  {\bf 89} (2002) 121601
[arXiv:hep-ph/0112260].
\bibitem{Hagiwara:fs}
K.~Hagiwara {\it et al.}  [Particle Data Group Collaboration],
Phys.\ Rev.\ D {\bf 66} (2002) 010001.
\bibitem{Aubert:2005ye}
B.~Aubert {\it et al.}  [BABAR Collaboration],
Phys.\ Rev.\ Lett.\  {\bf 95} (2005) 041802
[arXiv:hep-ex/0502032].
\bibitem{Allanach:2001kg}
B.~C.~Allanach,
Comput.\ Phys.\ Commun.\  {\bf 143} (2002) 305
[arXiv:hep-ph/0104145].
\bibitem{Kane:1998im}
G.~L.~Kane and S.~F.~King, Phys.\ Lett.\ B {\bf 451} (1999) 113
[arXiv:hep-ph/9810374].


\end{thebibliography}
\end{document}